\documentclass[accepted]{arXiv_OKD_20260227}
\usepackage[american]{babel}
\usepackage{natbib}
\usepackage{mathtools}
\usepackage{amssymb}
\usepackage{booktabs}
\usepackage{tikz}
\usepackage{algorithm}
\usepackage{algpseudocode}
\usepackage{rotating}
\usepackage{xcolor}
\usepackage{float}

\newcommand{\lag}{\ell}

\title{Operationalizing Longitudinal Causal Discovery Under Real-World Workflow Constraints}

\author[1]{\href{mailto:<okuda.tadahisa.3i@kyoto-u.ac.jp>}{Tadahisa Okuda}{}}
\author[2,3,4]{Shohei Shimizu}
\author[3]{Thong Pham}
\author[5]{Tatsuyoshi Ikenoue}
\author[1,6]{Shingo Fukuma}

\affil[1]{
    Kyoto University Graduate School of Medicine\\
    Kyoto, Japan
}
\affil[2]{
    SANKEN\\
    The University of Osaka\\
    Osaka, Japan
}
\affil[3]{
    Faculty of Data Science\\
    Shiga University\\
    Shiga, Japan
}
\affil[4]{%
    AIP\\
    RIKEN\\
    Tokyo, Japan
}
\affil[5]{
    Faculty of Medicine, University of Miyazaki\\
    Miyazaki, Japan
}
\affil[6]{
    Hiroshima University Graduate School of Biomedical and Health Sciences\\
    Hiroshima, Japan
}

\begin{document}
\maketitle

\begin{abstract}
    Causal discovery has achieved substantial theoretical progress, yet its deployment in large-scale longitudinal systems remains limited.
    A key obstacle is that operational data are generated under institutional workflows whose induced partial orders are rarely formalized, enlarging the admissible graph space in ways inconsistent with the recording process.
    We characterize a workflow-induced constraint class for longitudinal causal discovery that restricts the admissible directed acyclic graph space through protocol-derived structural masks and timeline-aligned indexing.
    Rather than introducing a new optimization algorithm, we show that explicitly encoding workflow-consistent partial orders reduces structural ambiguity, especially in mixed discrete--continuous panels where within-time orientation is weakly identified.
    The framework combines workflow-derived admissible-edge constraints, measurement-aligned time indexing and block structure, bootstrap-based uncertainty quantification for lagged total effects, and a dynamic representation supporting intervention queries.
    In a nationwide annual health screening cohort in Japan with 107,261 individuals and 429,044 person-years, workflow-constrained longitudinal LiNGAM yields temporally consistent within-time substructures and interpretable lagged total effects with explicit uncertainty.
    Sensitivity analyses using alternative exposure and body-composition definitions preserve the main qualitative patterns.
    We argue that formalizing workflow-derived constraint classes improves structural interpretability without relying on domain-specific edge specification, providing a reproducible bridge between operational workflows and longitudinal causal discovery under standard identifiability assumptions.
\end{abstract}

\section{Introduction: The Deployment Gap in Causal Discovery}\label{sec:intro}
Causal discovery has achieved substantial theoretical progress over the past two decades \citep{Spirtes93book,Peters17elements,Glymour19review}.
Under identifiable assumptions, structure-learning algorithms recover directed acyclic graphs (DAGs) that support causal interpretation and effect estimation.
In longitudinal settings, extensions of LiNGAM \citep{Shimizu06JMLR,Shimizu22book} and related methods exploit temporal ordering and non-Gaussianity to separate within-time and cross-time relations.

Despite these advances, a gap remains between identifiable theory and large-scale operational deployment \citep{Kotoku2020-ed,Uchida2022-nd,Fujita2023-nn}.
In many real-world panel systems, data are not generated under abstract time indices but under institutional workflows.
These workflows determine when variables are recorded, how exposures are assigned, what quantities summarize intervals, and how evaluation occurs.
When such workflow-induced partial orders are not formalized, the admissible DAG space implicitly includes structures inconsistent with the recording process, enlarging the search space and introducing avoidable structural ambiguity.

This issue is particularly pronounced in mixed discrete--continuous longitudinal panels.
Within-time orientation is often weakly identified, and minor preprocessing or indexing decisions can alter the set of Markov-equivalent orientations compatible with the data.
Standard forward-in-time constraints alone do not resolve this ambiguity, because ``time'' in recorded data may not coincide with ``causal time'' in the institutional process.

In this paper, we characterize a class of workflow-induced structural constraints for longitudinal causal discovery.
Rather than proposing a new optimization algorithm, we formalize how protocol-derived admissible-edge masks and timeline-aligned indexing restrict the effective search space of longitudinal DAGs.
Let $\mathcal{G}_\text{unconstrained}$ denote the set of DAGs admissible under standard forward-time constraints, and let $\mathcal{G}_\text{workflow}$ denote the subset further restricted by workflow-consistent partial orders.
By construction, $\mathcal{G}_\text{workflow} \subset \mathcal{G}_\text{unconstrained}$.
This restriction reduces structural ambiguity without imposing domain-specific medical directional assumptions, thereby modifying the set of graph structures that are identifiable in longitudinal causal discovery.

The proposed design layer rests on four principles:
\begin{enumerate}
    \item \textbf{Workflow-derived admissible-edge constraints.}
    Institutional ordering and recording properties are encoded as structural masks that restrict admissible edges independently of substantive domain claims.
    \item \textbf{Timeline-aligned indexing and block structure.}
    Modeled time points are aligned with evaluation schedules, and ordered within-time blocks reflect recording resolution, reducing orientation instability in mixed-type panels.
    \item \textbf{Uncertainty quantification for lagged total effects.}
    Subject-level bootstrap resampling provides uncertainty summaries directly tied to decision-relevant total effects in the constrained graph space.
    \item \textbf{Dynamic representation of the learned structure.}
    The estimated longitudinal system is represented as a linear dynamic model supporting forward and inverse intervention queries.
\end{enumerate}
Importantly, our contribution lies at the level of the admissible DAG class rather than the estimation algorithm: we do not alter the estimation routine of longitudinal LiNGAM, but instead redefine the graph class over which identifiability is assessed.
By aligning structural assumptions with workflow-induced partial orders, the framework reduces orientation non-identifiability arising from calendar--workflow mismatches while preserving standard LiNGAM assumptions (linearity, non-Gaussianity, and acyclicity within time).

As a population-scale stress test, we apply the framework to a nationwide annual health screening cohort in Japan ($107{,}261$ individuals; $429{,}044$ person-years; $15$ variables over four years).
Under workflow-constrained longitudinal LiNGAM, the learned structures exhibit temporally consistent within-time subgraphs and interpretable lagged total effects with quantified uncertainty.
Sensitivity analyses using alternative exposure and body-composition definitions preserve qualitative conclusions.

More broadly, longitudinal causal discovery in operational systems requires explicit separation between algorithmic foundations and constraint design.
When institutional workflows induce partial orders over variables and time, formalizing these constraint classes provides a reproducible mechanism for restricting the admissible DAG space, thereby improving structural interpretability under standard identifiability assumptions.

\section{Related Work}\label{sec:related}
This work builds on longitudinal causal discovery and focuses on the design layer required to translate structure learning into deployable decision-support systems.
We review prior work along four axes: longitudinal causal discovery, prior-knowledge constraints, uncertainty quantification, and decision-oriented use of learned causal models.

\paragraph{Longitudinal causal discovery.}
Causal discovery under identifiable assumptions has been extensively studied \citep{Spirtes93book,Peters17elements}.
LiNGAM and its variants, including DirectLiNGAM~\citep{Shimizu06JMLR,Shimizu11JMLR,Shimizu22book}, recover causal orderings under non-Gaussianity assumptions.
Longitudinal extensions~\citep{Hyva10JMLR,Kadowaki13MLSP} separate within-time and cross-time relations and typically restrict links to forward directions.
However, these approaches assume abstract time indexing and do not formalize workflow-induced partial orders or block-level admissibility constraints.
Our work follows longitudinal LiNGAM but introduces an explicit workflow-grounded restriction of the admissible DAG space for deployment settings.

\paragraph{Prior knowledge constraints.}
Many frameworks allow partial constraints to restrict the search space \citep{Spirtes93book,Peters17elements}, such as temporal precedence or prohibited edges.
We differ in formalizing constraints derived from recording protocols rather than domain-specific expert assumptions.
These protocol-level constraints encode when variables are measured and how interventions are assigned, preserving auditability and transferability.

\paragraph{Uncertainty quantification.}
Resampling and sensitivity analyses have been used to assess uncertainty in learned structures and effects \citep{Komatsu10ICANN}.
In deployment contexts, however, decision-relevant effect stability is central.
We therefore report bootstrap-based uncertainty for lagged total effects as primary outputs.

\paragraph{From graphs to decision systems.}
Causal models support counterfactual reasoning and policy analysis \citep{Pearl00book}, but causal discovery studies often stop at graph estimation.
We represent the learned longitudinal model as a dynamic intervention system enabling forward simulation and goal-oriented queries, and summarize stable within-time relations via recurring within-time subgraph patterns for operational monitoring.

\paragraph{Application context.}
Prior large-scale evaluations of the national health guidance program used eligibility thresholds to identify local effects via regression discontinuity design \citep{Fukuma2020-al}.
In contrast, we provide a workflow-grounded causal discovery pipeline that estimates longitudinal structure and system-level lagged effects with quantified uncertainty.

\section{Design Principles for Deployable Longitudinal Causal Discovery}\label{sec:design}
Deploying longitudinal causal discovery in operational systems requires more than selecting a structure-learning algorithm.
Real-world panel data are generated under fixed workflows, mix discrete and continuous variables, and are used to support decisions under uncertainty \citep{Krogsboll2019-am}.
We therefore distinguish the causal discovery method from a preceding design layer: explicit choices that translate workflow-generated data into algorithm-compatible inputs and decision-relevant outputs.
This section presents four principles that operationalize longitudinal causal discovery while preserving auditability and transferability.
An illustrative theoretical note on the resulting restriction of the admissible graph class is provided in Appendix~\ref{app:workflow_geom}.

\subsection{Workflow-Derived Structural Constraints}\label{sec:design-pk}
In operational longitudinal systems, the recording workflow fixes the order in which variables are observed and interventions are assigned.
In annual health screening programs, measurements are recorded, eligibility for guidance is determined, and downstream outcomes are observed at subsequent visits \citep{MHLW2010-wp,MHLW2024-wp}.
We encode these institutional orderings as workflow-derived structural constraints that restrict the admissible directed edges.

The prior knowledge used here is derived from (i) self-evident real-world information (e.g., age and sex are not modified by health guidance within the study window) and (ii) recording-protocol properties (what each variable represents, when it is recorded, and over what interval it summarizes).
We deliberately avoid expert-driven medical edge selection to reduce subjectivity and improve transferability.

Formally, let $\mathbf{X}^{(t)}$ denote the recorded variables at time $t$.
A prior-knowledge mask marks edges as allowed, forbidden, or unknown, and causal discovery is performed only within the resulting restricted graph class.
The mask enforces: (a) no time reversal, (b) forward cross-time links restricted to $t\!-\!1 \rightarrow t$, and (c) within-time admissibility consistent with the workflow.
Full details are given in Appendix~\ref{app:pk}.

\subsection{Timeline-Aligned Block Design for Mixed-Type Panels}\label{sec:design-block}
Real-world panels mix discrete indicators (e.g., intervention, medication, lifestyle) with continuous outcomes \citep{Curry2018-vh,Nishizawa2019-pb}.
Applying structure learning directly to such panels can yield unstable within-time orientations.
To mitigate this, we adopt workflow-aligned time indexing and block structure.

\paragraph{Time-point alignment.}
In annual screening systems, health guidance determined in year $y$ affects outcomes measured in year $y\!+\!1$.
We therefore define the modeled time point $t$ so that guidance corresponds to $y_{t}\!-\!1$ and outcomes to $y_{t}$.
This alignment respects the recording workflow and avoids introducing artificial cross-time links.

\paragraph{Block structure.}
Within each time point, variables are grouped into ordered blocks reflecting recording and measurement characteristics.
Guidance precedes questionnaire-based discrete variables, which precede continuous outcomes.
Directed relations are permitted only in directions consistent with this block order, while outcome-to-outcome relations remain unconstrained.
This block design reduces orientation ambiguity in mixed-type panels while preserving flexibility among outcomes.

\paragraph{Medication and lifestyle variables.}
We exclude within-time directed edges between medication status and lifestyle habits.
This exclusion is not a substantive medical claim, but a recording-based decision: both variables summarize behavior over the same interval, and the data do not identify their within-visit ordering.
Dependence is instead represented through cross-time links from $t\!-\!1$ to $t$.

\subsection{Bootstrap Uncertainty for Lagged Total Effects}\label{sec:design-uncertainty}
Deployment requires uncertainty summaries tied to decision questions.
We quantify uncertainty in lagged total causal effects using subject-level bootstrap resampling.
For each replicate ($B=1000$), individuals are resampled, the constrained model is refit, and total effects are computed for lags $\lag \in \{0,1,2\}$.
Uncertainty is summarized via empirical distributions and percentile confidence intervals.
These summaries align reported uncertainty with the quantities used in interpretation and planning.

\subsection{Learned Model as a Decision-relevant Representation}\label{sec:design-decision}
For deployment, the learned structure must be more than a static graph.
We recast the estimated longitudinal DAG and structural coefficients as a dynamic intervention model.

Let $\mathbf{B}^{(t,t)}$ denote within-time coefficients and $\mathbf{B}^{(t,t-1)}$ cross-time coefficients (restricted by prior knowledge).
Together with workflow-aligned indexing, these define a linear dynamic system mapping current states to future states under hypothetical interventions.

This representation enables (i) forward simulation (what-if analysis) and (ii) inverse target-setting queries that compute the upstream changes required to achieve specified downstream targets.
Binary variables are treated as switchable settings, while continuous outcomes remain primary targets.
Workflow-derived constraints, timeline alignment, bootstrap uncertainty, and decision-relevant representation together form a reusable procedure for deploying longitudinal causal discovery in workflow-generated panel data.

\section{Framework: Workflow-Constrained Longitudinal Causal Discovery}\label{sec:frame}
\subsection{Data and problem setup}\label{sec:frame_data}
We analyze a longitudinal panel constructed from annual health screening records over four consecutive years.
The final analytic cohort comprises 107{,}261 individuals (429{,}044 person-years).
At each time point, we define a 15-variable set consisting of:
(i) a binary indicator of program participation (\texttt{Health-guidance}),
(ii) five continuous health screening outcomes (body mass index, \texttt{BMI}; systolic blood pressure, \texttt{SBP}; diastolic blood pressure, \texttt{DBP}; hemoglobin A1c, \texttt{HbA1c}; low-density lipoprotein cholesterol, \texttt{LDL}),
(iii) three medication indicators (antihypertensive, \texttt{Drug-HT}; antidiabetic, \texttt{Drug-DM}; lipid-lowering, \texttt{Drug-LDL}),
(iv) three lifestyle indicators (smoking status, \texttt{Smoke}; exercise, \texttt{Exercise}; alcohol, \texttt{Alcohol}),
(v) age (\texttt{Age}) and sex (\texttt{Sex}), and
(vi) a categorical variable, \texttt{Check\_num.}, representing the number of health screening attendances during the three years preceding the first measurement year in the panel (2017--2019).

We use a 15-variable set that matches the routinely recorded fields available at every annual visit in this operational system.
This choice preserves comparability with the prior population-scale evaluation of the program \citep{Fukuma2020-al} and keeps the stress test grounded in recording structure and data availability.
The variable set is treated as a fixed design input rather than a contribution of this paper; the proposed framework applies without change when the set is expanded or modified, provided that the recording protocol and time alignment are specified.

The health guidance variable is defined as follows:
``participation'' denotes individuals who received health guidance and completed the program through the final evaluation; ``non-participation'' denotes all others (see Appendix~\ref{app:intervention_def}).

Separately, an assignment indicator can be defined from the program's institutional eligibility rule, which uses waist circumference thresholds together with additional cardiometabolic risk-factor criteria and excludes individuals under pharmacologic treatment for relevant conditions \citep{MHLW2024-wp}.
As shown in Appendix~\ref{app:assignment_def}, this assignment mechanism corresponds to a rule-based (conditional) intervention determined by observable eligibility criteria.
In this sense, assignment represents an institutional intervention rule, whereas participation reflects realized exposure.
The assignment indicator is used in a sensitivity analysis (see Section~\ref{sec:frame_sens}).

To apply longitudinal causal discovery, the data are organized as a three-dimensional array
(\texttt{subjects} $\times$ \texttt{variables} $\times$ \texttt{time points}),
and we adopt workflow-aligned time indexing reflecting one-year institutional ordering
(i.e., participation is determined after screening and evaluated against outcomes at the subsequent annual visit).
Appendix~\ref{app:variables} summarizes the variable definitions and time-point construction.

\subsection{Workflow-constrained longitudinal LiNGAM}\label{sec:frame_model}
We apply LiNGAM \citep{Shimizu06JMLR}, 
which assumes linear structural equations, non-Gaussian independent errors, 
acyclicity within each time point, and no hidden common causes among the modeled endogenous variables.
The model permits directed cross-time links, thereby capturing lagged causal effects across consecutive time points.

In our implementation, variables at time point 0 are treated as observed initial conditions (no within-time causal discovery) and serve as exogenous inputs for estimating relations at later time points (Figure~\ref{fig:pk_concept}).
Accordingly, the no-hidden-confounding assumption applies to the structural relations among the remaining endogenous variables, conditional on these observed exogenous factors.

Let $\mathbf{x}^{(t)}$ denote the $p$-dimensional vector of endogenous variables at time point $t$.
We use $v^{(t)}$ for the scalar intervention indicator, $\mathbf{z}^{(t)}$ for observed inputs (medications, lifestyle habits, and demographics), $w^{(0)}$ for the scalar baseline covariate used for adjustment, and $\mathbf{e}^{(t)}$ for error terms.
In our application, $\mathbf{x}^{(t)}$ collects the five continuous health screening outcomes (\texttt{BMI}, \texttt{SBP}, \texttt{DBP}, \texttt{HbA1c}, \texttt{LDL}) and $v^{(t)}$ denotes \texttt{Health-guidance}.
We let $\mathbf{z}^{(t)}$ collect medication and lifestyle indicators together with demographics, and $w^{(0)}$ be the baseline-only covariate (\texttt{Check\_num.}).
We adopt a first-order longitudinal formulation for $t=1,2,3$ as
\begin{equation}
    \begin{aligned}
        \mathbf{x}^{(t)}
        &=\ \boldsymbol{\alpha}^{(t,t)} v^{(t)}
            + \mathbf{B}^{(t,t)} \mathbf{x}^{(t)}
            + \mathbf{B}^{(t,t-1)} \mathbf{x}^{(t-1)} \\
        &\quad
            +\ \mathbf{C}^{(t,t)} \mathbf{z}^{(t)}
            + \mathbf{C}^{(t,t-1)} \mathbf{z}^{(t-1)} \\
        &\quad
            +\ \mathbb{I}\{t\!=\!1\}\,\boldsymbol{\delta}^{(1,0)} w^{(0)}
            + \mathbf{e}^{(t)},
    \end{aligned}
    \label{eq:model}
\end{equation}
where $\boldsymbol{\alpha}^{(t,t)}$ and $\boldsymbol{\delta}^{(1,0)}$ are $p$-dimensional coefficient vectors; $\mathbf{B}^{(t,t)}$ and $\mathbf{C}^{(t,t)}$ are coefficient matrices for within-time directed relations; and $\mathbf{B}^{(t,t-1)}$ and $\mathbf{C}^{(t,t-1)}$ are coefficient matrices for one-year lagged relations.
Here $\mathbb{I}\{t\!=\!1\}=1$ if $t=1$ and $0$ otherwise.

As illustrated in Figure~\ref{fig:pk_concept}, direct links with longer delays (e.g., $t\!-\!2 \to t$) are constrained to zero to align with the annual workflow and control model complexity. 
Longer-term influences are therefore represented through multi-step paths and summarized via total effects.
Time point 0 variables are treated as observed initial conditions and are not subject to within-time structure learning.
A baseline-only covariate (\texttt{Check\_num.}) enters only through $\mathbb{I}\{t\!=\!1\}\,\boldsymbol{\delta}^{(1,0)} w^{(0)}$.
We further impose workflow-derived prior knowledge, restricting structure learning to the endogenous variables in $\mathbf{x}^{(t)}$.
Variables in $\mathbf{z}^{(t)}$ and $w^{(0)}$ are not subject to causal discovery; rather, they are incorporated as observed exogenous inputs and serve as adjustment variables to control for confounding when estimating within-year causal relations among variables in $\mathbf{x}^{(t)}$ via DirectLiNGAM \citep{Shimizu11JMLR}.
The search space is restricted by workflow-derived prior knowledge based on the recording protocol (timing and nature of each variable) and basic data-structural facts, rather than medical edge selection.
Appendix~\ref{app:pk} details how these constraints are incorporated into estimation.

\subsection{Estimation and uncertainty quantification}\label{sec:frame_est}
Estimation follows a DirectLiNGAM-style iterative procedure under the workflow-derived constraints, alternating regression and independence assessment to infer a causal ordering and estimate coefficients.
Uncertainty in lagged total effects is quantified via subject-level bootstrap resampling with $B=1000$ replicates.
For each replicate, individuals are resampled with replacement, the constrained model is refit, and the propagating influences are computed to total effects through the fitted linear system.
Uncertainty is summarized using empirical bootstrap distributions and percentile confidence intervals, aligning reported variability with decision-relevant quantities.
A compact pseudo-code summary of the full workflow-constrained causal discovery procedure (including bootstrap uncertainty and total-effect computation) is provided in Appendix~\ref{app:algo_summary}.

\subsection{Sensitivity specifications}\label{sec:frame_sens}
We conduct two sensitivity analyses.
First, we replace BMI with alternative body-composition measures (waist circumference or body weight) and rerun the full pipeline without modifying other components.
Second, we replace the program participation indicator with the assignment indicator and rerun the pipeline under the same workflow-derived constraints and estimation procedure.
These analyses assess robustness to (i) alternative body-composition measures \citep{Helajarvi2014-oa,Vatcheva2016-qw,Tsushita2018-yc} and (ii) alternative exposure definitions (program participation versus rule-based assignment).

\section{Population-Scale Stress Test}\label{sec:res}
We report results from a population-scale application of the proposed workflow-constrained longitudinal causal discovery procedure to annual health screening data.
We focus on (i) lagged total effects of health guidance with bootstrap uncertainty summaries and (ii) consistent within-time subgraph structures among health screening outcomes that support compact interpretation under workflow-derived constraints.

\begin{table*}[t]
    \centering
    \caption{Total effects of \texttt{Health-guidance}(2020) on subsequent annual health screening outcomes measured in 2021--2023 (model lags 0--2; the first to third post-guidance annual visits).
    Each entry reports the point estimate and the 95\% percentile confidence interval (in parentheses) from subject-level bootstrap resampling ($B=1000$) under the workflow-derived prior-knowledge constraints.
    Total effects aggregate all directed paths in the fitted longitudinal system.}
    \label{tab:total_effects_2020}
    \begin{adjustbox}{max width=\textwidth}
        \normalsize
        \setlength{\tabcolsep}{10pt}
        \begin{tabular}{lccc}
            \toprule
            Outcome & 2021 (lag 0) & 2022 (lag 1) & 2023 (lag 2) \\
            \midrule
            BMI   & $-0.129$ [$-0.165,\ -0.094$] & $-0.067$ [$-0.109,\ -0.029$] & $-0.031$ [$-0.076,\ 0.014$] \\
            SBP   & $-0.737$ [$-1.112,\ -0.358$] & $-0.117$ [$-0.543,\ 0.290$]  & $0.203$  [$-0.250,\ 0.630$] \\
            DBP   & $-0.185$ [$-0.450,\ 0.080$]  & $0.305$  [$0.011,\ 0.591$]   & $0.531$  [$0.207,\ 0.837$] \\
            HbA1c & $-0.005$ [$-0.014,\ 0.005$]  & $-0.007$ [$-0.017,\ 0.005$]  & $0.002$  [$-0.010,\ 0.015$] \\
            LDL   & $-0.258$ [$-0.928,\ 0.439$]  & $0.086$  [$-0.683,\ 0.845$]  & $0.348$  [$-0.472,\ 1.175$] \\
            \bottomrule
        \end{tabular}
        \end{adjustbox}
    \begin{minipage}{\textwidth}
        \smaller
        \raggedright
        \textbf{Notes.} BMI: body mass index; SBP/DBP: systolic/diastolic blood pressure; HbA1c: hemoglobin A1c; LDL: low-density lipoprotein cholesterol.
    \end{minipage}
\end{table*}

\subsection{Lagged total effects of health guidance}\label{sec:res_te}
Table~\ref{tab:total_effects_2020} reports lagged total effects of health guidance (2020) on subsequent health screening outcomes (2021--2023), together with 95\% percentile confidence intervals from subject-level bootstrap resampling ($B=1000$).

The total effect on BMI is negative at lag~0 and remains negative at lag~1, with a smaller magnitude at longer lags.
For SBP, the lag~0 total effect is negative with an interval excluding zero, whereas uncertainty increases at longer lags.
For DBP, the lag~0 interval includes zero, while positive effects appear at lag~1 and lag~2.
For HbA1c and LDL, intervals include zero across the reported lags.

These summaries reflect combined direct and indirect pathways in the fitted longitudinal model under the workflow-derived constraints.
Because exposure is defined by program participation, the reported effects capture the influence of receiving and completing health guidance within the observed operational process.
DBP estimates tend to become more positive at later lags, which can reflect propagation through mediated cross-time pathways in the constrained linear dynamic system, although this tendency is less stable across sensitivity specifications (see Appendix~\ref{app:te_sens}).

\subsection{Bootstrap distributions and uncertainty patterns}\label{sec:res_boot}
\begin{figure*}[t]
    \centering
    \includegraphics[width=\textwidth]{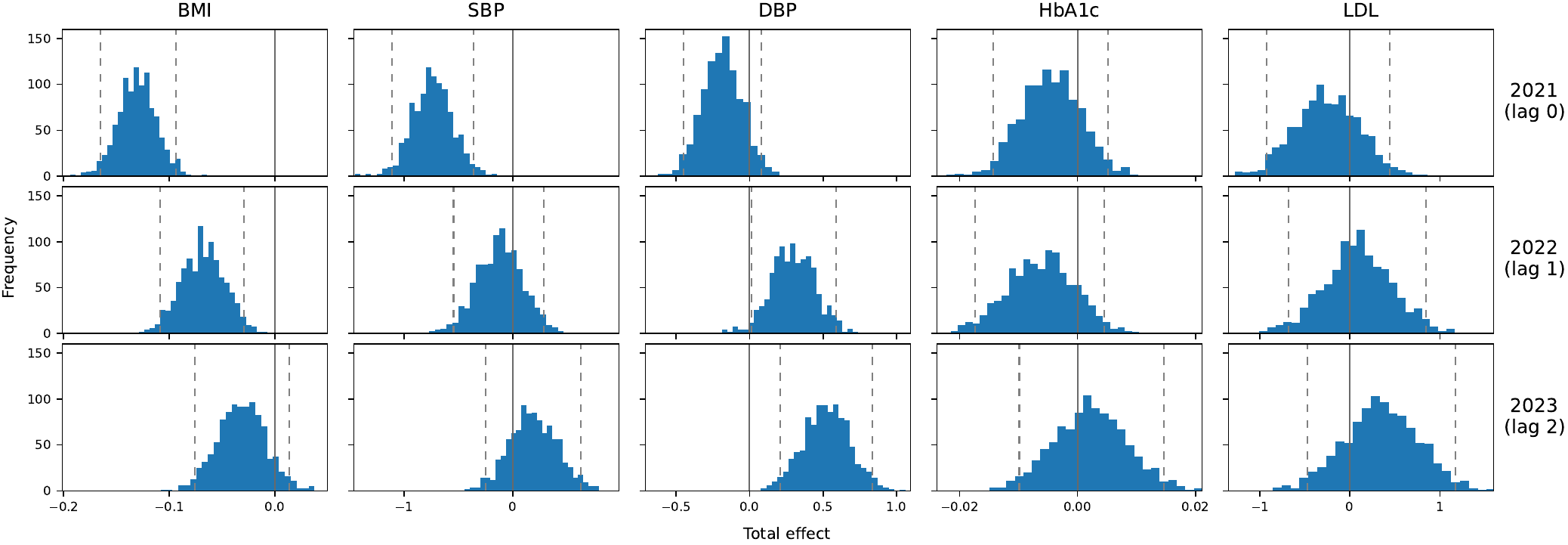}
    \caption{Bootstrap distributions of lagged total effects from health guidance in 2020 to outcomes measured in 2021--2023 (lags 0--2), under the workflow-derived constraints ($B=1000$).
    Panels are arranged in a $3\times 5$ grid: rows correspond to lag and columns to outcomes (BMI, SBP, DBP, HbA1c, LDL).
    Dashed vertical lines indicate the 95\% bootstrap percentile interval, and solid vertical lines mark zero.}
    \label{fig:te_hist_2020}
    \begin{minipage}{\textwidth}
        \raggedright
        \smaller
        \textbf{Notes.} BMI: body mass index; SBP/DBP: systolic/diastolic blood pressure; HbA1c: hemoglobin A1c; LDL: low-density lipoprotein cholesterol.
    \end{minipage}
\end{figure*}

Figure~\ref{fig:te_hist_2020} visualizes bootstrap distributions of total effects from \texttt{Health-guidance}(2020) to outcomes measured in 2021--2023.
Beyond the 95\% interval endpoints, the histograms show how uncertainty differs by outcome and by horizon through changes in spread and tail behavior.
Consistent with Table~\ref{tab:total_effects_2020}, BMI and SBP exhibit tighter distributions in the first post-guidance year with negative centers, whereas several outcomes at later years show wider distributions with substantial mass near zero.
DBP shows a tendency toward more positive draws at later years.

\subsection{Recurring Within-Time Graph Substructures}\label{sec:res_pattern}
\begin{figure}[t]
    \centering
    \includegraphics[width=\columnwidth]{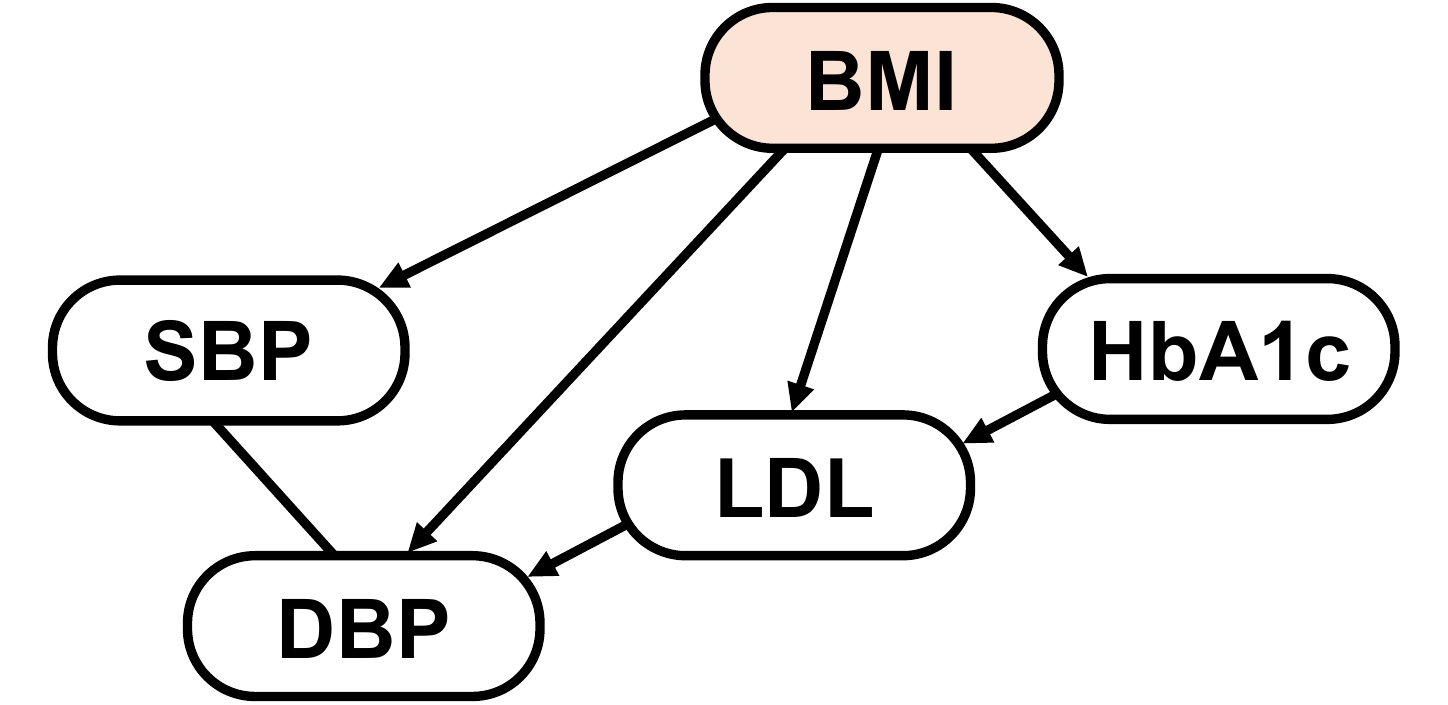}
    \caption{Compact recurring subgraph (motif) summarizing within-time relations among the five continuous health screening outcomes.
    Directed edges indicate directions that are consistent across time points 1--3 under the workflow-derived constraints.
    The undirected SBP--DBP link indicates a recurring adjacency whose direction varies across time points.
    The motif is a descriptive summary of recurrent within-time structure, not the full longitudinal graph.}
    \label{fig:motif_outcomes}
    \raggedright
    \smaller
    \textbf{Notes.} BMI: body mass index; SBP/DBP: systolic/diastolic blood pressure; HbA1c: hemoglobin A1c; LDL: low-density lipoprotein cholesterol.
\end{figure}

The learned longitudinal graphs are large, yet they show recurring within-time adjacency patterns across time points.
Figure~\ref{fig:motif_outcomes} summarizes this pattern as a compact motif over the five continuous health screening outcomes.
Directed edges indicate directions that are consistent across time points, whereas the undirected SBP--DBP connection denotes a recurring adjacency with time-varying direction.

We use the motif as an interpretable summary of multivariate within-time dependencies, while maintaining quantitative emphasis on lagged total effects (Table~\ref{tab:total_effects_2020}) and their bootstrap uncertainty (Figure~\ref{fig:te_hist_2020}).
The motif is a descriptive abstraction rather than a complete representation of the full longitudinal DAG.

\subsection{Sensitivity results}\label{sec:res_sens}
We stress-test the main total-effect conclusions under three alternative specifications: (i) replacing the program participation indicator with the rule-based assignment indicator, (ii) replacing BMI with waist circumference, and (iii) replacing BMI with body weight.
Across these variants, adiposity-related measures (e.g., BMI, waist circumference, and body weight) show the most stable short-horizon reductions at the first post-guidance annual visit.
For SBP, a short-horizon reduction is apparent under the participation-based exposure and body-composition variants, whereas assignment-based estimates are less pronounced and more uncertain.
At longer horizons, effects generally attenuate and uncertainty widens.
Full results are reported in Appendix~\ref{app:te_tables} and Appendix~\ref{app:te_hist_sens}.

\section{Generalization Beyond Healthcare}\label{sec:gen}
Although demonstrated in a healthcare setting, the proposed design layer relies on workflow-level properties rather than domain-specific medical assumptions.
More generally, the framework applies to longitudinal systems in which (i) institutional workflows induce partial orders over variables and time, (ii) heterogeneous variable types coexist within time slices, and (iii) structure-learning outputs are required to support repeated operational use.
We do not claim empirical validation beyond healthcare; the point is that the required inputs are recording- and workflow-level features that recur across deployed longitudinal settings.

\section{Discussion: From Algorithms to Infrastructure}\label{sec:dis}
This paper narrows the deployment gap in longitudinal causal discovery by specifying an explicit design layer that enables existing methods to operate under real-world workflow constraints.
Across a nationwide longitudinal health screening cohort, the combination of workflow-derived structural constraints, timeline-aligned block design, and bootstrap-based uncertainty summaries yielded interpretable system-level results (lagged total effects with empirical intervals) together with a compact summary of consistent within-time graph substructures.

\subsection{Why the design layer is necessary}\label{sec:dis_need}
Longitudinal causal discovery in operational settings differs from benchmark settings in one central respect: the data are generated by institutional workflows.
In our study setting, measurements are recorded, program participation occurs (or not), and subsequent measurements are obtained according to a fixed annual schedule.
When workflow-induced ordering is ignored, mixed-type observational data are likely to admit multiple Markov-equivalent orientations; restricting the admissible space reduces avoidable ambiguity.
The results illustrate the practical value of this approach.

First, the lagged total effects provide decision-relevant summaries that integrate direct and indirect pathways without privileging individual edges.
The strongest and most reliable signal appears at lag~0 for BMI (and, under the participation-based exposure, SBP), with attenuation at longer lags, while uncertainty broadens for several outcomes as lag increases (Table~\ref{tab:total_effects_2020}).

Second, the learned graphs exhibit recurring within-time graph substructures across time points.
We summarize this recurring pattern as a compact motif over the five continuous outcomes (Figure~\ref{fig:motif_outcomes}): directed edges indicate directions that are consistent across time points, whereas the undirected SBP--DBP connection denotes a recurring adjacency with time-varying direction.
This motif is not presented as a definitive mechanistic claim; rather, it provides an interpretable abstraction of recurrent multivariate substructures that supports (i) communication to non-technical stakeholders and (ii) downstream uses such as monitoring structural drift and performing simulator-based queries for forward prediction and inverse target-setting.

From a program-evaluation perspective, the most stable signal in our system-level summaries is concentrated in body-weight-related measures.
In Table~\ref{tab:total_effects_2020}, the lag-0 total effect on BMI is negative with a confidence interval excluding zero, whereas later lags show attenuation with broader uncertainty (see also Figure~\ref{fig:te_hist_2020}).
This pattern is consistent with the program's operational intent---guidance primarily targets behaviors and weight control first, with downstream changes propagating through the broader longitudinal system---while we avoid attributing the result to any single directed edge.
Importantly, the same qualitative message persists under our sensitivity specifications that replace BMI with alternative adiposity measures and that use the eligibility-based assignment indicator in place of observed participation (see Appendix~\ref{app:te_sens}).
In that limited sense, the short-term reduction and later attenuation we observe are also broadly aligned with the time profile reported in assignment-based evaluations by \citet{Fukuma2020-al}.
Our contribution complements that line of work by adding a workflow-grounded design layer and interpretable system-level summaries (graphs, motifs, and simulator queries) rather than focusing on a single local causal contrast.

Historically, as cohort studies made population risk factors measurable, regression models became part of the standard analytic infrastructure \citep{Mahmood2014-qa}.
Analogously, as operational systems demand causal conclusions while limiting reliance on subjective edge specification, workflow-grounded causal discovery can serve as infrastructure for deployment-ready longitudinal analysis.

\subsection{Limitations}\label{sec:dis_lim}
This study has several limitations.
Longitudinal LiNGAM relies on linearity, non-Gaussian independent errors, acyclicity within time points, and the absence of hidden common causes among endogenous variables conditional on observed baseline covariates.
As in most observational settings, these assumptions are not fully testable, and unmeasured confounding cannot be excluded.
We therefore interpret the recovered structure and lagged total effects as workflow-grounded, assumption-dependent summaries useful for generating transparent hypotheses and decision-relevant sensitivity checks rather than definitive causal truth.

Future work could replace longitudinal LiNGAM with extensions that relax the no-hidden-confounding assumption, such as methods that allow more flexible structural forms \citep{Maeda20AISTATS,Salehkaleybar2020learning,Pham26AISTATS}.
The proposed design layer is compatible with such extensions, and incorporating them would strengthen robustness to unmeasured confounding while preserving workflow-derived constraints.

Second, mixed discrete and continuous variables are modeled within a primarily linear framework.
The block design reduces orientation instability but does not eliminate possible model misspecification.

Third, the primary exposure is defined as program participation (completion of health guidance).
Although operationally meaningful, this definition may introduce selection effects related to adherence and follow-up.
To mitigate interpretation risk, we also analyze the rule-based assignment indicator while keeping the remainder of the pipeline fixed; the main qualitative patterns are preserved, although some endpoints (e.g., SBP) become less pronounced and more uncertain.

Fourth, bootstrap resampling captures sampling variability within the specified modeling pipeline but does not reflect all uncertainty sources, including systematic measurement error, violations of modeling assumptions, or temporal shifts in the underlying population.

Finally, the panel is annual, and cross-time links are restricted to one-year lags.
Because measurements are annual, causal changes occurring within a year cannot be resolved, and direct multi-year delays cannot be distinguished from cumulative one-year effects. Extending the analysis to longer horizons would additionally require modeling potential structural changes over time, such as policy shifts or changes in measurement procedures.

\subsection{Deployment considerations}\label{sec:dis_deploy}
A central aim of this work is to make causal discovery outputs usable in practice.
First, outputs must remain stable under routine re-training as new cohorts accumulate.
Reporting lagged total effects with bootstrap intervals together with a compact recurring subgraph facilitates monitoring for structural drift more effectively than relying on a single dense graph.

Second, the learned model is represented as a decision-relevant dynamic system.
Given a selected current variable and a counterfactual value, forward simulation yields predicted future outcomes.
Conversely, inverse target-setting computes the upstream change required to achieve a specified downstream outcome at a chosen horizon.
Binary variables, including program participation, are treated as switchable inputs, while continuous health outcomes remain primary targets.

The design layer supports deployment by (i) defining admissible relations under workflow constraints, (ii) stabilizing estimation under mixed types, and (iii) aligning uncertainty reporting with decision-relevant quantities.
Full deployment additionally requires user interfaces, governance structures, and monitoring for dataset shift.
We have implemented a practitioner-facing prototype supporting forward and inverse queries, and pilot stakeholder evaluation is underway.
Detailed information is provided in Appendix~\ref{app:prototype}.

\section{Conclusion}\label{sec:con}
We presented a framework that makes explicit the design layer required for longitudinal causal discovery under workflow-generated panel data.
Rather than modifying the optimization procedure itself, our contribution lies in formally characterizing how workflow-induced structural constraints restrict the admissible DAG class and thereby alter the effective search space over which identifiability is assessed.

By encoding institutional partial orders and recording-resolution properties as admissible-edge masks, we induce a strict subset of the unconstrained longitudinal graph space.
This restriction reduces structural ambiguity that arises when calendar time is misaligned with workflow time, particularly in mixed discrete--continuous panels where within-time orientation is weakly identified.
In this sense, the contribution lies at the level of the admissible graph space rather than the estimation algorithm, modifying the set of candidate causal structures in a reproducible and auditable manner without relying on domain-specific edge specification.

In a population-scale longitudinal cohort, the proposed constraint formalization yielded (i) temporally consistent within-time substructures and (ii) interpretable lagged total effects with bootstrap uncertainty summaries.
These outputs arise not from additional modeling complexity, but from aligning structural assumptions with the data-generating workflow.

More broadly, longitudinal causal discovery in operational systems benefits from explicitly separating algorithmic foundations (identifiability theory and estimation procedures) from constraint design.
When institutional workflows induce partial orders over variables and time, making those constraints explicit can improve structural interpretability while preserving identifiability under standard assumptions.
We argue that formalizing workflow-derived constraint classes constitutes a necessary step toward reproducible and deployment-ready longitudinal causal discovery.

\bibliography{bib_OKD.bib}

\newpage
\onecolumn
\title{Operationalizing Longitudinal Causal Discovery Under Real-World Workflow Constraints (Supplementary Material)}
\maketitle
\appendix
\numberwithin{figure}{section}
\numberwithin{table}{section}
\numberwithin{equation}{section}

\section{Data, Variables, and Operational Definitions}\label{app:data_def}
\subsection{Variable set and panel layout}\label{app:variables}
\begin{figure}[htp]
    \centering
    \includegraphics[width=\linewidth]{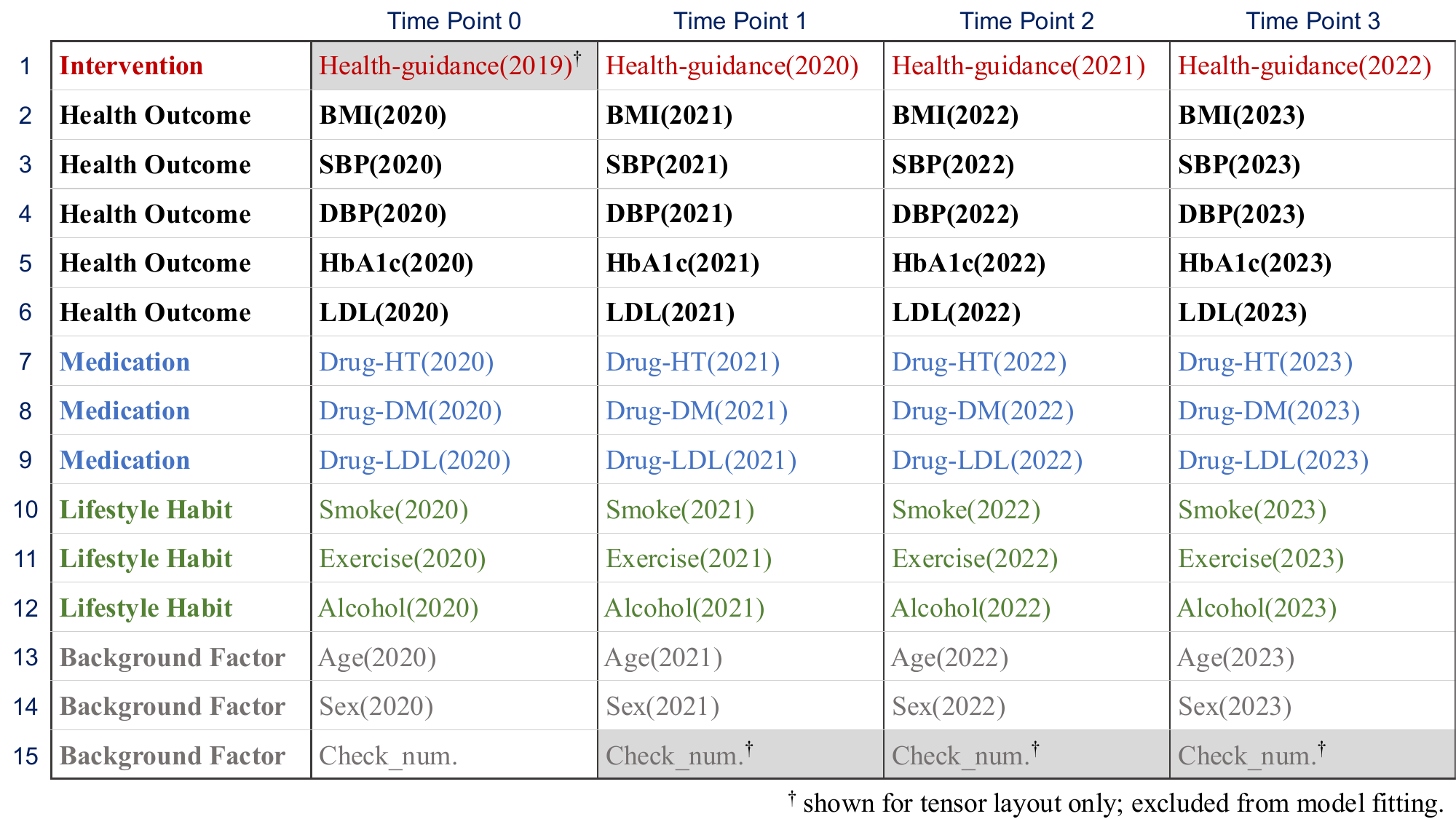}
    \caption{Variable set and time-point alignment used in this study.
    Each time point contains 15 variables: health guidance, five continuous health screening outcomes (BMI, SBP, DBP, HbA1c, LDL), three medication indicators, three lifestyle indicators, demographics (Age, Sex), and \texttt{Check\_num.} (attendance history in the three years preceding the measurement year).
    Cells marked with \textsuperscript{\ensuremath{\dagger}} are shown only to illustrate the fixed tensor layout across time points and are excluded from model fitting.}
    \label{fig:var_set}
\end{figure}

We analyze a longitudinal panel constructed from annual health screening records over four consecutive years.
The final analytic cohort comprises 107{,}261 individuals (429{,}044 person-years).
At each time point, we define a 15-variable set consisting of:
\begin{itemize}
    \item\textbf{A binary indicator of intervention:}
    \texttt{Health-guidance}.
    \item\textbf{Five continuous health screening outcomes:}
    body mass index, \texttt{BMI}; systolic blood pressure, \texttt{SBP}; diastolic blood pressure, \texttt{DBP}; hemoglobin A1c, \texttt{HbA1c}; low-density lipoprotein cholesterol, \texttt{LDL}.
    \item\textbf{Three medication indicators:}
    antihypertensive, \texttt{Drug-HT}; antidiabetic, \texttt{Drug-DM}; lipid-lowering, \texttt{Drug-LDL}.
    \item\textbf{Three lifestyle indicators:}
    smoking status, \texttt{Smoke}; exercise habits, \texttt{Exercise}; alcohol use, \texttt{Alcohol}.
    \item\textbf{Demographics:}
    age, \texttt{Age}; sex, \texttt{Sex}.
    \item\textbf{A categorical attendance-history variable:}
    \texttt{Check\_num.}, representing the number of health screening attendances during the three years preceding the measurement year (2017--2019).
\end{itemize}
Figure~\ref{fig:var_set} summarizes the variables and their alignment across time points.
Entries marked with \textsuperscript{\ensuremath{\dagger}} are shown for layout illustration only and are excluded from model fitting and effect reporting.
Accordingly, \texttt{Health-guidance}(2019) at time point 0 is displayed only to illustrate the fixed tensor layout across time points.
Likewise, \texttt{Check\_num.} is a baseline-only covariate used only for adjustment; it is displayed at later time points in the figure solely to illustrate the fixed tensor layout.

\subsection{Health guidance indicator (intervention definition)}\label{app:intervention_def}
The primary exposure is an intervention indicator derived from administrative program records.
We define \texttt{Health-guidance}=1 for individuals who received health guidance and completed the program through the final evaluation, and \texttt{Health-guidance}=0 for all others.

\subsection{Assignment indicator (sensitivity-analysis exposure)}\label{app:assignment_def}
\begin{figure}[htp]
    \centering
    \includegraphics[width=\linewidth]{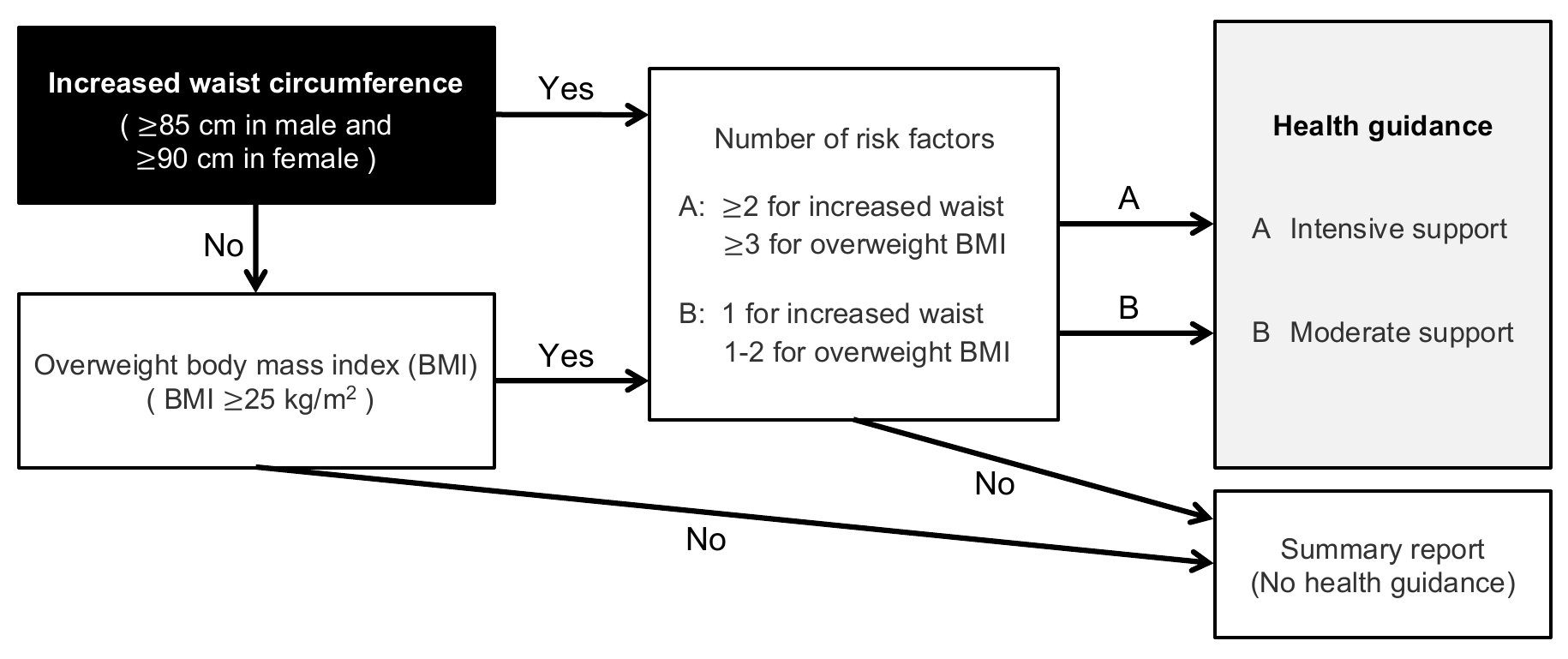}
    \caption{Schematic of the health guidance assignment logic used to construct the assignment indicator for sensitivity analysis.
    The selection proceeds from waist circumference screening to assessment of risk factor status, with medication-treated individuals excluded from eligibility (details are simplified for exposition).}
    \label{fig:in_select}
\end{figure}

Separately, an assignment indicator can be defined from the program's selection rule.
The rule includes an initial screening step based on waist circumference, followed by assessment of risk factor status.
It excludes individuals receiving pharmacologic treatment for blood pressure, glycemia, or lipids from eligibility.
Figure~\ref{fig:in_select} provides a schematic summary of this selection logic.
This assignment indicator is used only in sensitivity analysis; other components of the pipeline remain unchanged.

\section{Prior Knowledge and Block-Structured Design}\label{app:pk}
\subsection{Mapping between recorded variables and model notation}\label{app:role_mapping}
Table~\ref{tab:role_mapping} maps the recorded variables to the notation used in Equation~\ref{eq:model}.
Only $\mathbf{x}^{(t)}$ is subject to within-time structure learning; variables at time point 0 are treated as baseline covariates (adjustment only) and are not used for within-time causal discovery.

\begin{table}[htp]
    \centering
    \caption{Role mapping between recorded variables and the notation in Equation~\ref{eq:model}.}
    \label{tab:role_mapping}
    \begin{adjustbox}{max width=\textwidth}
        \normalsize
        \setlength{\tabcolsep}{10pt}
        \begin{tabular}{@{}lll@{}}
            \toprule
            Symbol & Contents in this study & Subject to causal discovery? \\
            \midrule
            $v^{(t)}$ & \texttt{Health-guidance}$(t)$ & No (observed input) \\
            $\mathbf{x}^{(t)}$ & \texttt{BMI}, \texttt{SBP}, \texttt{DBP}, \texttt{HbA1c}, \texttt{LDL} $(t)$ & Yes (within-time among outcomes) \\
            $\mathbf{z}^{(t)}$ & \texttt{Drug-HT}, \texttt{Drug-DM}, \texttt{Drug-LDL}, \texttt{Smoke}, \texttt{Exercise}, \texttt{Alcohol}, \texttt{Age}, \texttt{Sex} $(t)$ & No (observed inputs) \\
            $w^{(0)}$ & \texttt{Check\_num.} (baseline-only) & No (adjustment only) \\
            \bottomrule
        \end{tabular}
    \end{adjustbox}
\end{table}

\subsection{What the prior knowledge represents}\label{app:pk_source}
We incorporate prior knowledge as a structural mask that constrains the admissible edge set before performing longitudinal causal discovery.
The purpose is to encode operationally determined ordering and data-structural constraints so that the learned graphs remain consistent with the way the records are generated.

The prior knowledge in this work is derived from two sources:
\begin{enumerate}
    \item \textbf{Generic invariances and bookkeeping constraints} that hold irrespective of medical theory,
    e.g., \texttt{Sex} is time-invariant and \texttt{Age} evolves deterministically over time and is not affected by other recorded variables.
    \item \textbf{Recording-protocol constraints} implied by the annual health screening workflow,
    including the fact that measurements and questionnaire items are recorded at each annual visit and that health guidance status is recorded based on administrative program information.
\end{enumerate}
We do not encode medical causal claims (e.g., physiology-driven directions among biomarkers) as prior knowledge; within the admissible space, those relations are learned from data.

\subsection{prior-knowledge mask specification}\label{app:pk_mask}
The prior-knowledge mask is implemented as a time- and lag-indexed constraint tensor.
For within-time relations ($\mathrm{lag}=0$), entries take values in $\{-1,0,1\}$ to indicate unknown, forbidden, or allowed directions, following the convention used in DirectLiNGAM-style prior-knowledge masks.
For cross-time relations ($\mathrm{lag}>0$), entries are binary $\{0,1\}$ indicating whether an edge from time $t-\mathrm{lag}$ to time $t$ is allowed.

Cross-time links are restricted to one-year lag ($t\!-\!1 \to t$), and direct links with longer delays ($t\!-\!2\to t$, $t\!-\!3\to t$) are set to zero to control complexity and to align with the annual workflow.
Under the measurement-year indexing, \texttt{Health-guidance} at time point $t$ summarizes health guidance delivered in the interval since the previous annual visit and is aligned with measurements recorded at the current visit.
Therefore, a within-time edge from \texttt{Health-guidance}$(t)$ to a measurement at time point $t$ already represents the one-interval influence on that measurement.
To avoid encoding the same one-interval exposure window twice, we set direct cross-time edges from \texttt{Health-guidance}$(t\!-\!1)$ to measurements at time point $t$ to zero and represent delayed influence through intermediate variables.

\subsection{Block-structured within-time ordering}\label{app:block_design}
As described in Section~\ref{sec:design} and Figure~\ref{fig:var_set}, time points are indexed by the measurement year.
Under this indexing, \texttt{Health-guidance} at time point $t$ represents guidance delivered in the interval after the previous annual visit and before the current visit, so within-time links from \texttt{Health-guidance} to measurements at $t$ correspond to one-year influence.
Within each time point, we use a block-structured ordering that is consistent with how variables are recorded and used in the workflow.
In brief, the within-time block design allows the following dependencies at time point $t$.
\begin{itemize}
    \item \texttt{Health-guidance}$(t)$ has no instantaneous parents.
    \item Medication indicators and lifestyle indicators at $t$ may depend on \texttt{Health-guidance}$(t)$ and background factors (e.g., \texttt{Age}$(t)$, \texttt{Sex}).
    \item Each health screening outcome at $t$ may depend on \texttt{Health-guidance}$(t)$, background factors, and the medication/lifestyle indicators at $t$.
    \item Within-time directions among the five continuous outcomes are left to be learned (subject to acyclicity and the mask).
\end{itemize}

We treat the within-time component as describing dependencies among variables that are observed with the same temporal support at an annual visit.
The five health screening outcomes (BMI, SBP, DBP, HbA1c, LDL) are point measurements taken at the visit and therefore share a common time anchor, so within-time relations among them serve as an interpretable abstraction of contemporaneous physiological mechanisms at the visit.
In contrast, medication and lifestyle indicators are recorded as questionnaire/record summaries over the interval since the previous annual visit; the data do not resolve the within-interval ordering of changes between medication use and lifestyle habits within the same year.
Allowing within-time directed links between these two domains would therefore impose an artificial temporal ordering that the records do not support.
Accordingly, dependence between medication and lifestyle is modeled at the annual resolution through cross-time links from $t$$-$$1$ to $t$.
Within-time links from medication and lifestyle to the health screening outcomes remain admissible under the prior knowledge.
The separation follows recording resolution and does not encode a medical assumption that medication and lifestyle do not interact.

\subsection{Baseline-only variable}\label{app:checknum_pk}
The attendance-history variable \texttt{Check\_num.} is defined using the three years preceding the first measurement year and is treated as a baseline-only variable.
Accordingly, it is allowed to affect variables at the first modeled time point via $t\!-\!1\to t$ links (from baseline to the first annual visit), and it is isolated thereafter (i.e., it has no parents and is not used as a parent at later time points).

\subsection{prior-knowledge concept diagram}\label{app:pk_fig}
\begin{figure}[htp]
    \centering
    \includegraphics[width=\linewidth]{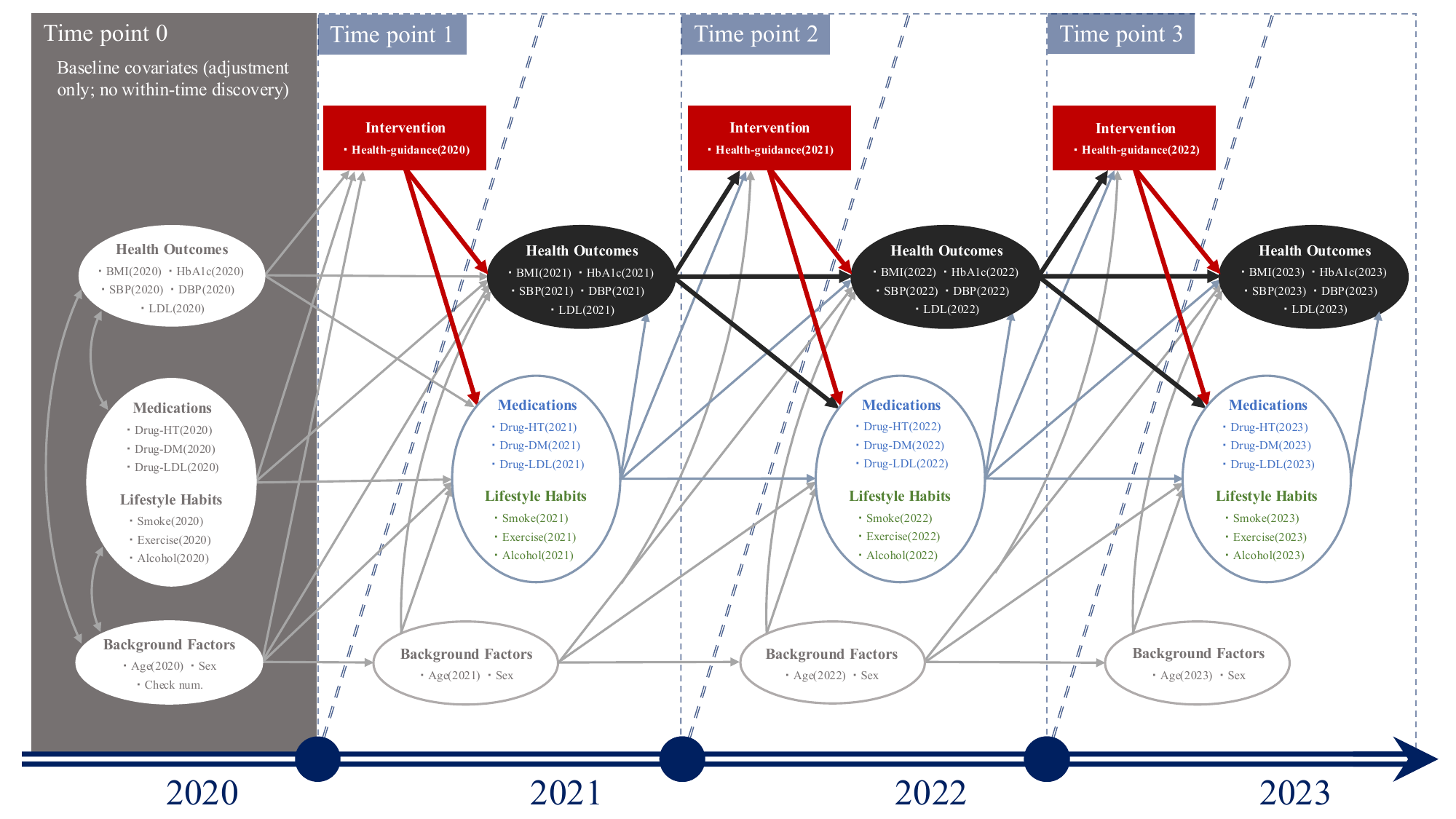}
    \caption{Conceptual diagram of the workflow-derived prior knowledge.
    The prior knowledge encodes (i) time ordering across annual visits, (ii) within-time block ordering consistent with the recording protocol, and (iii) admissible cross-time links restricted to one-year lag ($t\!-\!1\to t$).
    The prior knowledge is derived from generic invariances and recording-protocol constraints, not from medical causal assumptions.}
    \label{fig:pk_concept}
    \begin{minipage}{\linewidth}
        \raggedright
        \smaller
        \textbf{Notes.} Bidirectional links at time point 0 indicate baseline covariates used for adjustment only and not subject to within-time structure learning.
    \end{minipage}
\end{figure}

Figure~\ref{fig:pk_concept} summarizes the workflow-derived prior knowledge used in this study, including the time-point alignment, admissible edge directions, and the within-time block ordering.

\section{Supplementary theoretical note and algorithmic summary}\label{app:supp_theory_algo}
\subsection{Illustration of workflow-induced constraint classes}\label{app:workflow_geom}
By construction, $\mathcal{G}_\text{workflow} \subset \mathcal{G}_\text{unconstrained}$.
This inclusion can rule out orientations that are Markov-equivalent from conditional independences alone but violate the workflow-induced partial order.
In mixed discrete--continuous panels, where within-time orientation is often weakly identified, restricting the admissible edge space decreases structural ambiguity without imposing domain-specific medical directionality.
The contribution of the design layer lies at the level of the admissible graph space rather than the estimation algorithm: it modifies the effective candidate space over which identifiability is assessed in practice.

\subsection{Algorithmic summary of the workflow-constrained pipeline}\label{app:algo_summary}
Algorithm~\ref{alg:pipeline} summarizes the end-to-end pipeline used in this study, from applying the workflow-derived prior-knowledge mask (Appendix~\ref{app:pk}) to fitting the constrained longitudinal model and reporting bootstrap uncertainty for lag-specific total effects.
The purpose of this summary is to make the procedure reproducible at the level of inputs and steps; implementation details are provided in the main text and Appendix~\ref{app:pk} (mask design) and Appendix~\ref{app:te_sens} (additional results).

\begin{algorithm}[htp]
    \caption{Workflow-constrained longitudinal causal discovery with bootstrap total effects}\label{alg:pipeline}
    \begin{algorithmic}[1]
        \Require Annual panel tensor $X$; workflow-derived prior-knowledge mask \textit{PK} (Appendix~\ref{app:pk}); bootstrap size $B$
        \Ensure Coefficient blocks; lag-specific total effects with percentile intervals; within-time motif summary
            \State Align time points by measurement year and define within-time blocks consistent with the recording protocol
            \State Apply the workflow-derived mask \textit{PK} to restrict within-time and one-year cross-time links
            \State Fit workflow-constrained longitudinal LiNGAM to obtain within-time and one-year cross-time coefficient blocks
            \State Compute lag-specific total effects from the fitted longitudinal system for an anchor year and subsequent measurement years
            \For{$b=1,\dots,B$}
                \State Resample subjects with replacement and reconstruct the full panel
                \State Refit the constrained longitudinal model under the same mask \textit{PK}
                \State Recompute lag-specific total effects and store bootstrap draws
            \EndFor
            \State Report percentile confidence intervals from bootstrap distributions
            \State Summarize recurring within-time structure among continuous outcomes as a motif (Figure~\ref{fig:motif_outcomes}); see Appendix~\ref{app:motif_def} for extraction rules
    \end{algorithmic}
\end{algorithm}

\section{Additional total-effect results and sensitivity analyses}\label{app:te_sens}
\paragraph{Notation and abbreviations.}
Throughout this appendix, HG denotes health guidance; BMI, body mass index; Weight, body weight; Waist, waist circumference; SBP/DBP, systolic/diastolic blood pressure; HbA1c, hemoglobin A1c; LDL, low-density lipoprotein cholesterol.

\subsection{Bootstrap uncertainty quantification}\label{app:bootstrap}
Uncertainty is quantified by subject-level bootstrap resampling with $B=1000$ replicates.
In each replicate, we resample individuals with replacement, reconstruct the full four-year panel for the resampled cohort, refit the workflow-constrained longitudinal model under the same prior-knowledge mask, and recompute the reported total effects.
We report 95\% confidence intervals based on the empirical bootstrap distribution (percentile intervals).

\subsection{Lag-0 total effects by measurement year}\label{app:te_lag0_by_year}
Table~\ref{tab:sup_te_lag0_by_year} reports lag-0 (within-time under the prior knowledge) total effects by measurement year.
Time points are indexed by the measurement year, and \texttt{Health-guidance} recorded at time point $t$ represents guidance delivered during the interval between the previous and current annual visits.
Accordingly, lag~0 corresponds to the first post-guidance annual visit under this indexing (i.e., outcomes measured at year $t$ after guidance delivered in the preceding interval).
Each entry reports the point estimate and the 95\% percentile confidence interval (in parentheses) from subject-level bootstrap resampling ($B=1000$) under the workflow-derived prior-knowledge constraints.

\begin{table*}[htp]
    \centering
    \caption{Model-lag-0 total effects by measurement year.
    Under the measurement-year indexing, \texttt{Health-guidance} at time point $t$ summarizes guidance delivered during the interval preceding the visit, so model lag 0 corresponds to effects observed at the next annual visit.
    Each entry reports the point estimate and the 95\% percentile confidence interval (in parentheses) from subject-level bootstrap resampling ($B=1000$) under the workflow-derived prior-knowledge constraints.}
    \label{tab:sup_te_lag0_by_year}
    \begin{adjustbox}{max width=\textwidth}
        \normalsize
        \setlength{\tabcolsep}{10pt}
        \begin{tabular}{lccc}
            \toprule
            Outcome &
            \shortstack{HG 2020\\$\rightarrow$ 2021} &
            \shortstack{HG 2021\\$\rightarrow$ 2022} &
            \shortstack{HG 2022\\$\rightarrow$ 2023} \\
            \midrule
            BMI   & $-0.129$ [$-0.165,\ -0.094$] & $-0.082$ [$-0.114,\ -0.051$] & $-0.077$ [$-0.107,\ -0.046$] \\
            SBP   & $-0.737$ [$-1.112,\ -0.358$] & $-0.038$ [$-0.367,\ 0.306$]  & $-0.246$ [$-0.602,\ 0.109$] \\
            DBP   & $-0.185$ [$-0.450,\ 0.080$]  & $0.364$  [$0.127,\ 0.611$]   & $0.290$  [$0.053,\ 0.530$] \\
            HbA1c & $-0.005$ [$-0.014,\ 0.005$]  & $-0.011$ [$-0.019,\ -0.002$] & $-0.004$ [$-0.012,\ 0.004$] \\
            LDL   & $-0.258$ [$-0.928,\ 0.439$]  & $-0.391$ [$-1.035,\ 0.270$]  & $0.420$  [$-0.217,\ 1.052$] \\
            \bottomrule
        \end{tabular}
    \end{adjustbox}
\end{table*}

\subsection{Lag-specific total effects anchored at guidance year 2020}\label{app:te_tables}
Tables~\ref{tab:sup_te_assignment2020}--\ref{tab:sup_te_weight2020} report lag-specific total effects anchored at \texttt{Health-guidance}(2020) under the primary specification and sensitivity settings.
Under measurement-year indexing, outcomes measured in 2021, 2022, and 2023 correspond to model lags 0, 1, and 2, respectively (i.e., the first, second, and third post-guidance annual visits for guidance year 2020).
Total effects aggregate all directed paths in the fitted longitudinal system.
Negative values correspond to reductions in the outcome scale (e.g., lower BMI, blood pressure, HbA1c, or LDL).
The assignment indicator is constructed from the program’s selection rule, summarized in Appendix~A.3 (Figure~\ref{fig:in_select}).

\begin{table*}[htp]
    \centering
    \caption{Sensitivity analysis: total effects of the \textbf{assignment indicator} (eligibility-based) in 2020 on subsequent annual health screening outcomes (2021--2023; assignment defined by Figure~\ref{fig:in_select}).
    Each entry reports the point estimate and the 95\% percentile confidence interval (in parentheses) from subject-level bootstrap resampling ($B=1000$) under the workflow-derived prior-knowledge constraints.
    Total effects aggregate all directed paths in the fitted longitudinal system.}
    \label{tab:sup_te_assignment2020}
    \begin{adjustbox}{max width=\textwidth}
        \normalsize
        \setlength{\tabcolsep}{10pt}
        \begin{tabular}{lccc}
            \toprule
            Outcome & 2021 (lag 0) & 2022 (lag 1) & 2023 (lag 2) \\
            \midrule
            BMI   & $-0.050$ [$-0.067,\ -0.031$] & $-0.069$ [$-0.091,\ -0.046$] & $-0.059$ [$-0.084,\ -0.035$] \\
            SBP   & $0.064$  [$-0.131,\ 0.267$]  & $0.116$  [$-0.083,\ 0.321$]  & $0.055$  [$-0.163,\ 0.269$] \\
            DBP   & $0.638$  [$0.502,\ 0.783$]   & $0.737$  [$0.582,\ 0.876$]   & $0.655$  [$0.496,\ 0.814$] \\
            HbA1c & $-0.003$ [$-0.010,\ 0.004$]  & $-0.006$ [$-0.013,\ 0.002$]  & $0.002$  [$-0.006,\ 0.010$] \\
            LDL   & $0.416$  [$0.031,\ 0.777$]   & $-0.101$ [$-0.505,\ 0.291$]  & $-0.166$ [$-0.594,\ 0.243$] \\
            \bottomrule
        \end{tabular}
    \end{adjustbox}
\end{table*}

\begin{table*}[htp]
    \centering
    \caption{Sensitivity analysis: total effects of \texttt{Health-guidance}(2020) on subsequent annual health screening outcomes (2021--2023) when replacing BMI with \textbf{waist circumference}.
    Each entry reports the point estimate and the 95\% percentile confidence interval (in parentheses) from subject-level bootstrap resampling ($B=1000$) under the workflow-derived prior-knowledge constraints.
    Total effects aggregate all directed paths in the fitted longitudinal system.}
    \label{tab:sup_te_weight2020}
    \begin{adjustbox}{max width=\textwidth}
        \normalsize
        \setlength{\tabcolsep}{10pt}
        \begin{tabular}{lccc}
            \toprule
            Outcome & 2021 (lag 0) & 2022 (lag 1) & 2023 (lag 2) \\
            \midrule
            Waist & $-0.325$ [$-0.460,\ -0.199$] & $-0.181$ [$-0.333,\ -0.042$] & $-0.025$ [$-0.194,\ 0.124$] \\
            SBP   & $-0.668$ [$-1.042,\ -0.279$] & $-0.069$ [$-0.490,\ 0.344$]  & $0.268$  [$-0.178,\ 0.694$] \\
            DBP   & $-0.164$ [$-0.431,\ 0.102$]  & $0.330$  [$0.037,\ 0.614$]   & $0.555$  [$0.235,\ 0.854$] \\
            HbA1c & $-0.004$ [$-0.013,\ 0.006$]  & $-0.004$ [$-0.016,\ 0.007$]  & $0.004$  [$-0.008,\ 0.017$] \\
            LDL   & $-0.218$ [$-0.890,\ 0.496$]  & $0.127$  [$-0.653,\ 0.889$]  & $0.370$  [$-0.466,\ 1.182$] \\
            \bottomrule
        \end{tabular}
    \end{adjustbox}
\end{table*}

\begin{table*}[htp]
    \centering
    \caption{Sensitivity analysis: total effects of \texttt{Health-guidance}(2020) on subsequent annual health screening outcomes (2021--2023) when replacing BMI with \textbf{body weight}.
    Each entry reports the point estimate and the 95\% percentile confidence interval (in parentheses) from subject-level bootstrap resampling ($B=1000$) under the workflow-derived prior-knowledge constraints.
    Total effects aggregate all directed paths in the fitted longitudinal system.}
    \label{tab:sup_te_weight2020}
    \begin{adjustbox}{max width=\textwidth}
        \normalsize
        \setlength{\tabcolsep}{10pt}
        \begin{tabular}{lccc}
            \toprule
            Outcome & 2021 (lag 0) & 2022 (lag 1) & 2023 (lag 2) \\
            \midrule
            Weight & $-0.403$ [$-0.505,\ -0.300$] & $-0.212$ [$-0.337,\ -0.102$] & $-0.111$ [$-0.247,\ 0.019$] \\
            SBP    & $-0.629$ [$-1.004,\ -0.248$] & $-0.017$ [$-0.426,\ 0.386$]  & $0.329$  [$-0.119,\ 0.756$] \\
            DBP    & $-0.170$ [$-0.434,\ 0.097$]  & $0.326$  [$0.035,\ 0.613$]   & $0.554$  [$0.236,\ 0.861$] \\
            HbA1c  & $-0.003$ [$-0.013,\ 0.007$]  & $-0.004$ [$-0.015,\ 0.008$]  & $0.005$  [$-0.007,\ 0.018$] \\
            LDL    & $-0.189$ [$-0.859,\ 0.502$]  & $0.165$  [$-0.607,\ 0.917$]  & $0.405$  [$-0.426,\ 1.204$] \\
            \bottomrule
        \end{tabular}
    \end{adjustbox}
\end{table*}

\subsection{Bootstrap distributions under sensitivity settings}\label{app:te_hist_sens}
\begin{figure}[htp]
    \centering
    \includegraphics[width=\linewidth]{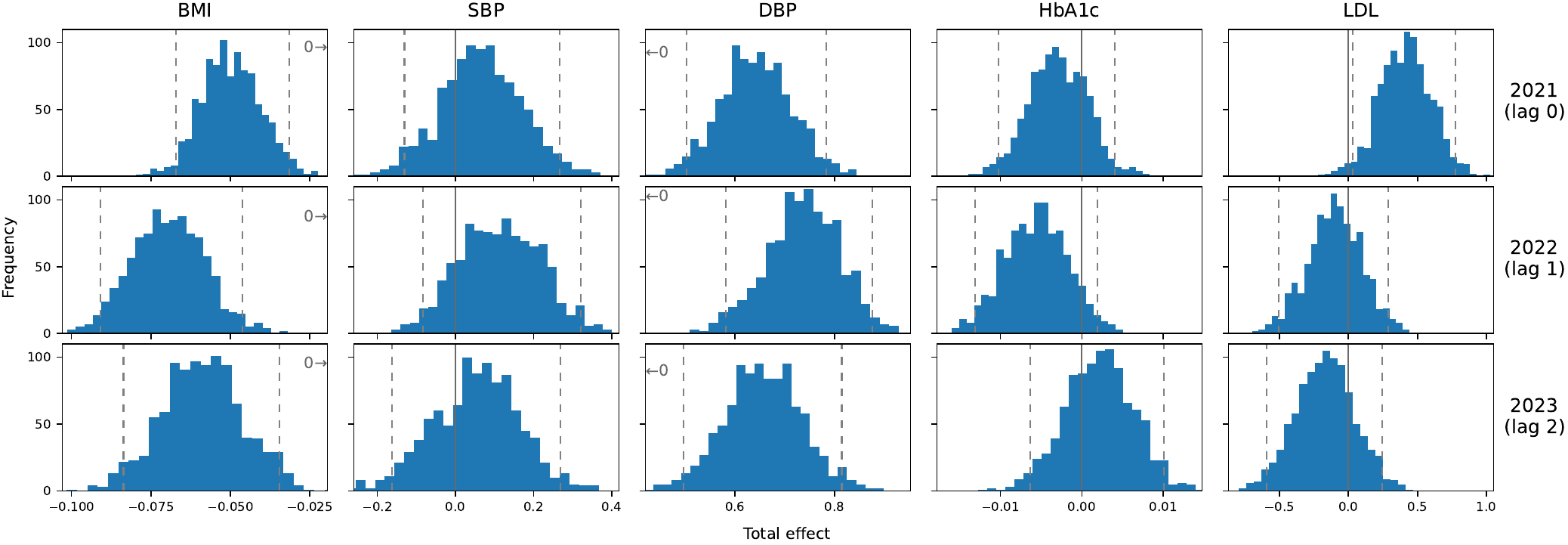}
    \caption{Sensitivity analysis (assignment indicator): bootstrap distributions of lag-specific total effects anchored at 2020. Rows correspond to outcome measurement years 2021--2023 (lags 0--2), and columns correspond to outcomes (BMI, SBP, DBP, HbA1c, LDL).
    Dashed vertical lines indicate the 95\% bootstrap percentile interval; the solid vertical lines mark zero when it lies within the plotted range (otherwise zero is indicated by an arrow).
    Distributions are computed by subject-level bootstrap resampling ($B=1000$) under the workflow-derived prior-knowledge constraints.}
    \label{fig:te_hist_assign_2020}
\end{figure}

\begin{figure}[htp]
    \centering
    \includegraphics[width=\linewidth]{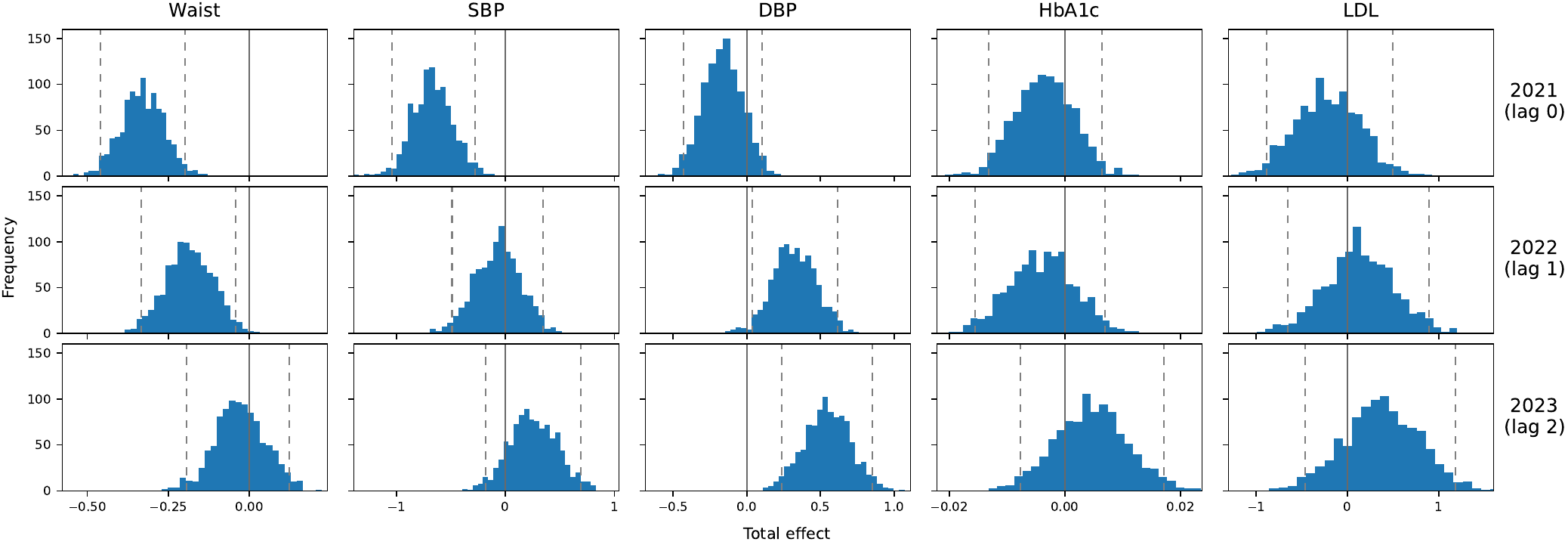}
    \caption{Sensitivity analysis (waist circumference): bootstrap distributions of lag-specific total effects anchored at 2020 when replacing BMI with waist circumference.
    Rows correspond to outcome measurement years 2021--2023 (lags 0--2), and columns correspond to outcomes (Waist, SBP, DBP, HbA1c, LDL).
    Dashed vertical lines indicate the 95\% bootstrap percentile interval, and solid vertical lines mark zero.
    Distributions are computed by subject-level bootstrap resampling ($B=1000$) under the workflow-derived prior-knowledge constraints.}
    \label{fig:te_hist_waist_2020}
\end{figure}

\begin{figure}[htp]
    \centering
    \includegraphics[width=\linewidth]{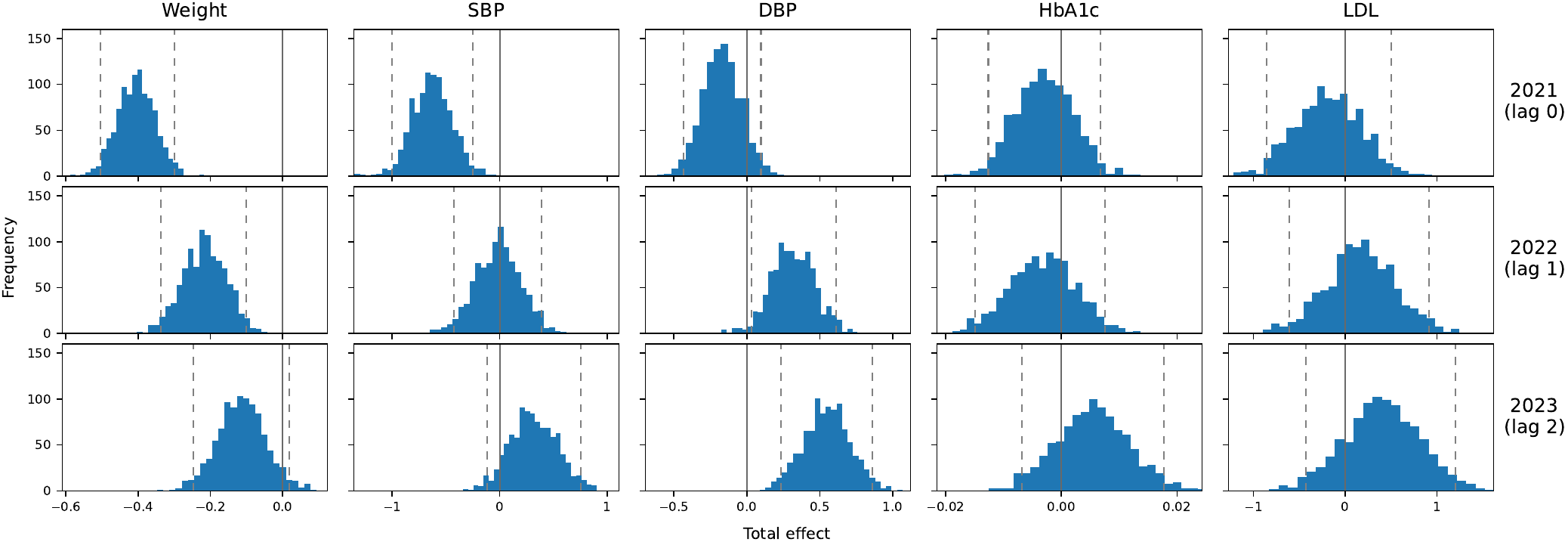}
    \caption{Sensitivity analysis (body weight): bootstrap distributions of lag-specific total effects anchored at 2020 when replacing BMI with body weight.
    Rows correspond to outcome measurement years 2021--2023 (lags 0--2), and columns correspond to outcomes (Weight, SBP, DBP, HbA1c, LDL).
    Dashed vertical lines indicate the 95\% bootstrap percentile interval, and solid vertical lines mark zero.
    Distributions are computed by subject-level bootstrap resampling ($B=1000$) under the workflow-derived prior-knowledge constraints.}
    \label{fig:te_hist_weight_2020}
\end{figure}

Figures~\ref{fig:te_hist_assign_2020}--\ref{fig:te_hist_weight_2020} visualize the bootstrap distributions of lag-specific total effects under the sensitivity settings.
Rows correspond to outcome measurement years 2021--2023 (lags 0--2 under measurement-year indexing), and columns correspond to the five continuous health screening outcomes.
These plots complement the tables by showing the shape and dispersion of uncertainty beyond percentile summaries.

\subsection{Notes on endpoint- and horizon-dependent variability}\label{app:te_notes}
Across outcomes, uncertainty generally increases at longer horizons, and some endpoints are more sensitive to alternative operational proxies than others.
In our sensitivity specifications (Appendix~\ref{app:te_sens}), adiposity-related measures (i.e., Waist, Weight) and SBP show similar short-horizon patterns, whereas DBP exhibits more horizon-dependent variability, including shifts in sign at later horizons.
We report this as an empirical feature of the workflow-constrained longitudinal system under alternative exposure and measurement definitions, rather than as a standalone mechanistic or clinical claim.

\section{Complete learned graph and motif extraction}\label{app:comp_graph}
\subsection{Full learned longitudinal graph}\label{app:full_graph}
\begin{sidewaysfigure*}[htp]
    \centering
    \includegraphics[width=\textheight]{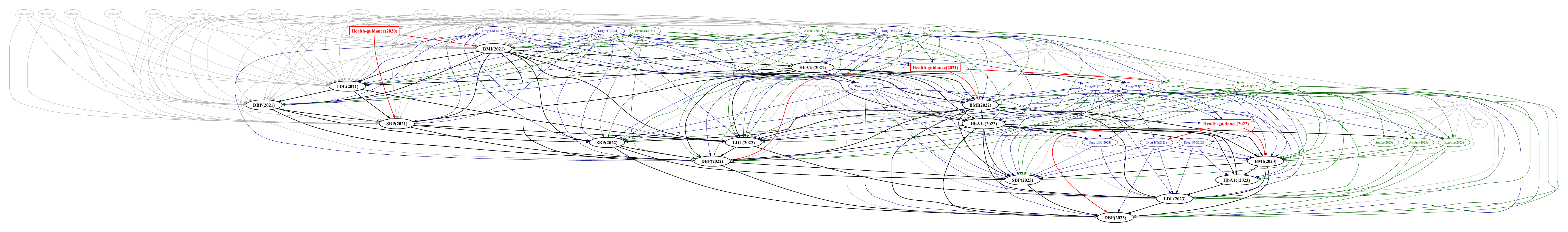}
    \caption{Full learned longitudinal directed graph (point estimate) under the primary specification.
    The graph is estimated using workflow-constrained longitudinal causal discovery with the prior-knowledge mask described in Appendix~\ref{app:pk}.
    Time points follow measurement-year indexing, and cross-time links are restricted to one-year lag ($t\!-\!1 \to t$).
    The figure shows the full structure from which the compact motif summary in the main text is extracted.
    Edges with $|\hat\beta|<0.01$ are omitted by the package's default visualization.}
    \label{fig:full_dag}
    \begin{minipage}{\textwidth}
        \raggedright
        \smaller
        \textbf{Notes.} \texttt{BMI}: body mass index; \texttt{SBP}/\texttt{DBP}: systolic/diastolic blood pressure; \texttt{HbA1c}: hemoglobin A1c; \texttt{LDL}: low-density lipoprotein cholesterol; \texttt{Drug-HT}: antihypertensive; \texttt{Drug-DM}: antidiabetic; \texttt{Drug-LDL}: lipid-lowering; \texttt{Smoke}: smoking status; \texttt{Exercise}: exercise habit; \texttt{Alcohol}: alcohol use; \texttt{Age}: age; \texttt{Sex}: sex; \texttt{Check\_num.}: representing the number of health screening attendances during the three years preceding the measurement year (2017--2019).
    \end{minipage}
\end{sidewaysfigure*}

\begin{figure}[htp]
    \centering
    \includegraphics[width=\linewidth,trim=1020 430 3070 100,clip]{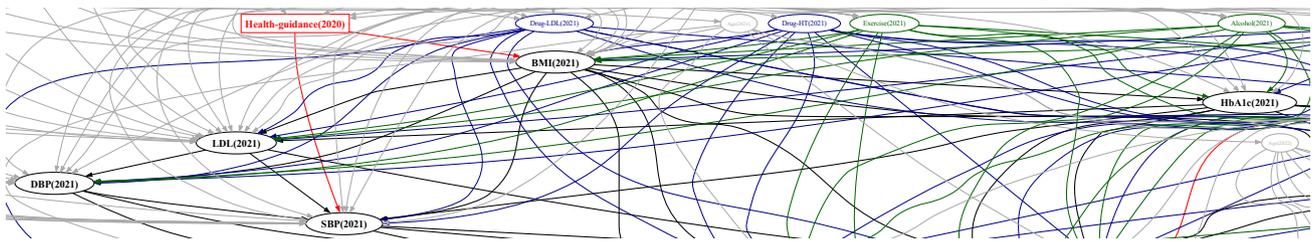}
    \caption{Zoomed view (crop) of Figure~\ref{fig:full_dag} for the first block (\texttt{Health-guidance}(2020) and outcomes measured in 2021).}
    \label{fig:full_dag_zoom_t1}
\end{figure}

\begin{figure}[htp]
    \centering
    \includegraphics[width=\linewidth,trim=2470 270 2350 260,clip]{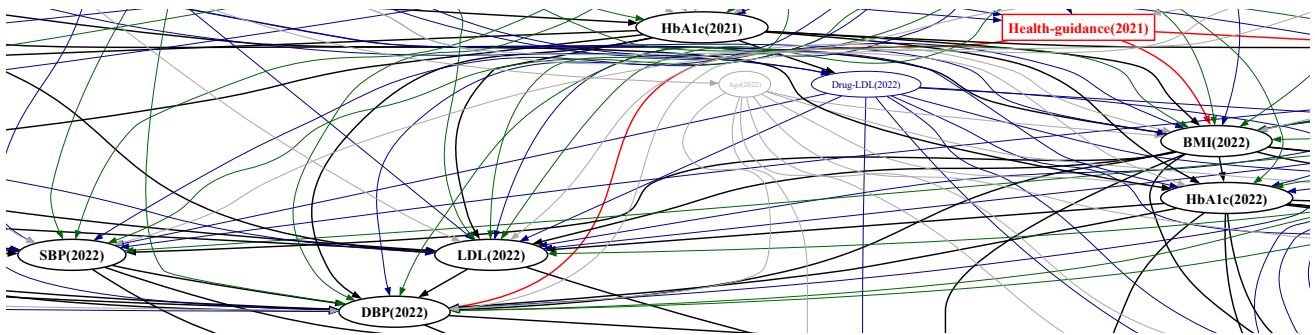}
    \caption{Zoomed view (crop) of Figure~\ref{fig:full_dag} for the second block (\texttt{Health-guidance}(2021) and outcomes measured in 2022).}
    \label{fig:full_dag_zoom_t2}
\end{figure}

\begin{figure}[htp]
    \centering
    \includegraphics[width=\linewidth,trim=4220 30 1180 490,clip]{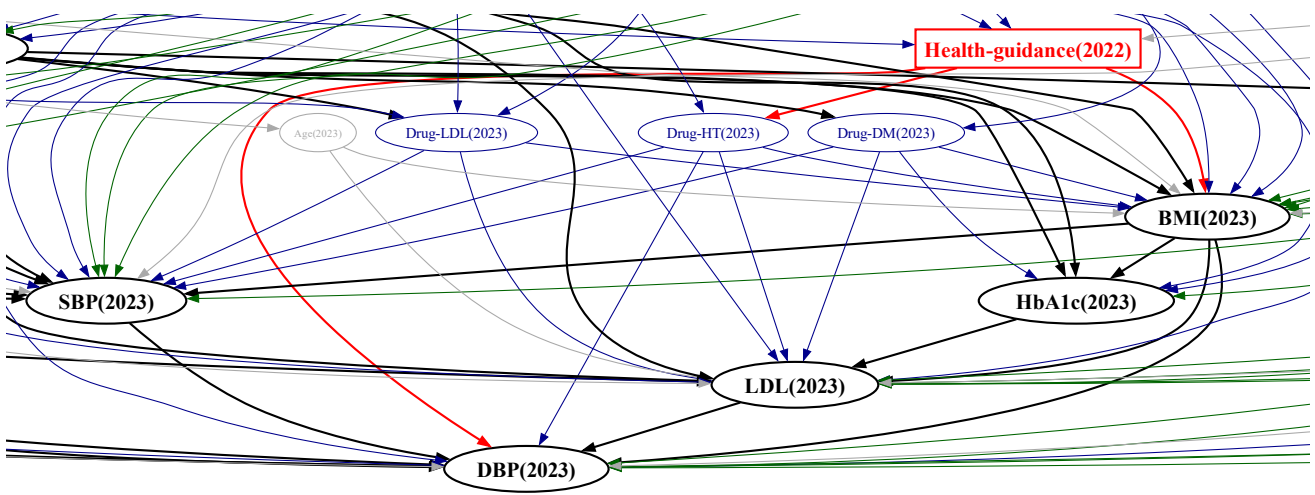}
    \caption{Zoomed view (crop) of Figure~\ref{fig:full_dag} for the third block (\texttt{Health-guidance}(2022) and outcomes measured in 2023).}
    \label{fig:full_dag_zoom_t3}
\end{figure}

Figure~\ref{fig:full_dag}--\ref{fig:full_dag_zoom_t3} present the full longitudinal directed graph learned under the primary specification (workflow-derived prior-knowledge mask, measurement-year indexing, and one-year cross-time links).
Nodes are arranged by time point and grouped by variable type (background, health guidance, medications, lifestyle indicators, and continuous outcomes).
Edges include within-time links and one-year cross-time links ($t\!-\!1 \to t$) that are admissible under the prior knowledge.


\subsection{Motif extraction rule and relation to the full longitudinal graph}\label{app:motif_def}
The motif in Figure~\ref{fig:motif_outcomes} summarizes within-time relations among the five continuous health screening outcomes that recur across all three time points under the workflow-derived prior knowledge.
Directed edges are included only when the corresponding within-time relation appears with a consistent direction across time points 1--3.
In addition, when an adjacency recurs across time points but its direction differs at each time point, we treat it as an undirected connection (SBP--DBP in our setting).
The motif is used as an interpretability device: it provides a compact summary of recurring local structure without presenting the full longitudinal graph or treating any single edge as a mechanistic claim.

\section{Deployment prototype: practitioner-facing simulator}\label{app:prototype}
\subsection{Role and intended use}\label{app:proto_role}
The learned longitudinal system is packaged into a practitioner-facing ``what-if'' simulator intended for operational use in preventive health guidance workflows.
The prototype is designed for scenario exploration and communication with practitioners, rather than for automated decision-making.
Outputs are computed from the fitted population-level model and are intended to support human judgment (human-in-the-loop; HITL) when evaluating plausible changes to modifiable factors and their downstream implications.

\subsection{User-facing functions}\label{app:proto_functions}
The interface provides two core functions.

\paragraph{Forward prediction (what-if).}
A user selects (i) a current-year variable to intervene on and (ii) a future-year target variable of interest, then inputs a hypothetical changed value for the selected current variable.
The simulator returns the model-implied expected value of the selected future target at a user-chosen horizon (one or two annual visits ahead), conditional on the baseline profile.

\paragraph{Goal seeking (target setting).}
A user selects (i) a future-year target variable and its desired value, then selects a current-year variable to adjust.
The simulator returns the model-implied value of the current variable required to achieve the desired future target (when a unique or stable solution exists under the fitted system).

Both functions support continuous and binary variables (e.g., medication and lifestyle indicators), and the displayed message template adapts to the selected variable types.

\subsection{Inputs and variable mapping}\label{app:proto_inputs}
Users first enter baseline inputs corresponding to the current annual visit.
Inputs align with the study variable set (continuous health screening outcomes, medication indicators, lifestyle indicators, and guidance status) and use the same coding as in model fitting.
The simulator restricts interactions to variables available in routine records, consistent with the paper's design-layer principle of operational availability.

\subsection{Computation from the fitted longitudinal system}\label{app:proto_compute}
The simulator evaluates queries using the fitted longitudinal structural model under the workflow-derived prior-knowledge constraints.
Forward prediction propagates the hypothetical current-year change through the system to obtain the implied expected value of the chosen future target.
Goal seeking solves the inverse problem for a selected pair of (current variable, future target) using the fitted equations. When the inverse is ill-posed or unstable, the interface reports that the query is not supported under the current configuration.

\paragraph{Lightweight deployment via precomputed block matrices.}
To ensure low-latency response in routine use, the prototype stores four precomputed block matrices covering all variable pairs and lag-specific blocks under a common specification: (i) point estimates of total effects, (ii) lower confidence bounds, (iii) upper confidence bounds, and (iv) a binary uncertainty indicator used for guardrail messaging.
At runtime, the Excel/VBA layer performs only lookup operations (row--column crosspoints) and basic arithmetic for formatting, avoiding expensive computations on the client side.
This separation between offline estimation and online lookup improves responsiveness and reduces operational fragility.

\subsection{Decision messaging and uncertainty guardrails}\label{app:proto_guardrails}
The prototype includes conservative messaging rules to avoid overconfident or misleading outputs.

\paragraph{Uncertainty-aware ``no detectable effect'' messaging.}
For a given query, when the 95\% bootstrap confidence interval of the corresponding total effect includes zero, the interface suppresses a numerical recommendation and instead displays a ``no statistically detectable effect'' message.
This rule matches the bootstrap uncertainty quantification used throughout the paper and is intended to reduce misinterpretation by end users.

\paragraph{Input validation and plausibility checks.}
The interface applies basic validation to prevent obvious data-entry errors (e.g., physiologically implausible values).
More formal out-of-support checks (e.g., thresholds based on empirical quantiles or other criteria) are under active development. They are treated as a conservative safeguard rather than as a definitive clinical judgment.

\paragraph{HITL positioning.}
These guardrails are designed to complement, not replace, practitioner expertise: the simulator provides model-based scenario summaries while leaving final interpretation and action to human users.

\subsection{Screenshots with English callouts}\label{app:proto_screens}
The operational prototype targets Japanese practitioners and is implemented with a Japanese user interface.
For readability in this paper, figures overlay concise English callouts directly on the screenshots (the \LaTeX\ source contains no non-ASCII user interface text).

\begin{figure}[H]
    \centering
    \includegraphics[width=\linewidth]{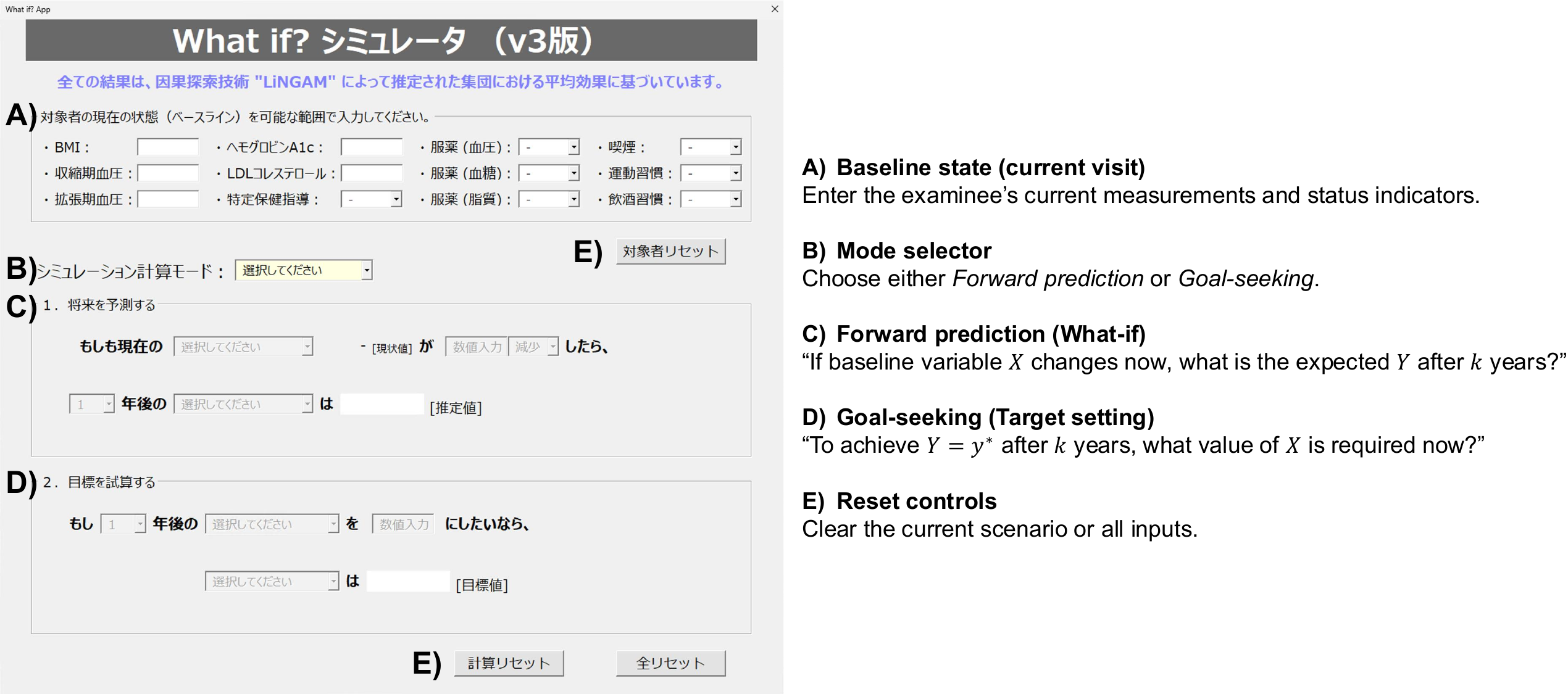}
    \caption{Overview of the practitioner-facing simulator (user interface in Japanese; English callouts added for the paper).}
    \label{fig:proto_overview}
\end{figure}

\begin{figure}[H]
    \centering
    \includegraphics[width=\linewidth]{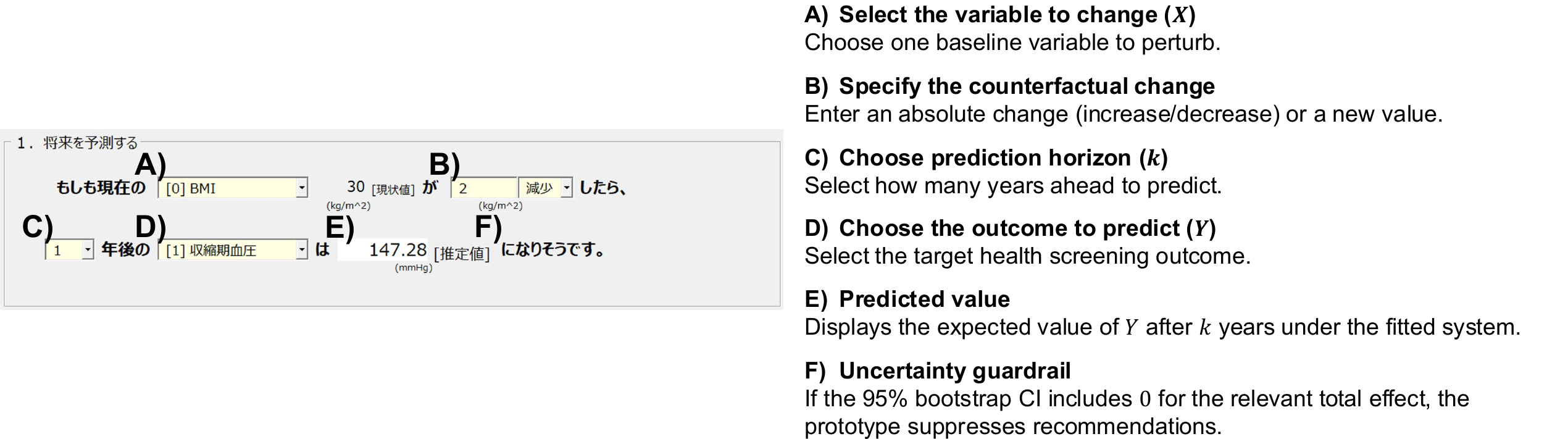}
    \caption{Forward prediction (What-if): changing a baseline variable X and predicting outcome Y after k years under the fitted longitudinal system.}
    \label{fig:proto_forward}
\end{figure}

\begin{figure}[H]
    \centering
    \includegraphics[width=\linewidth]{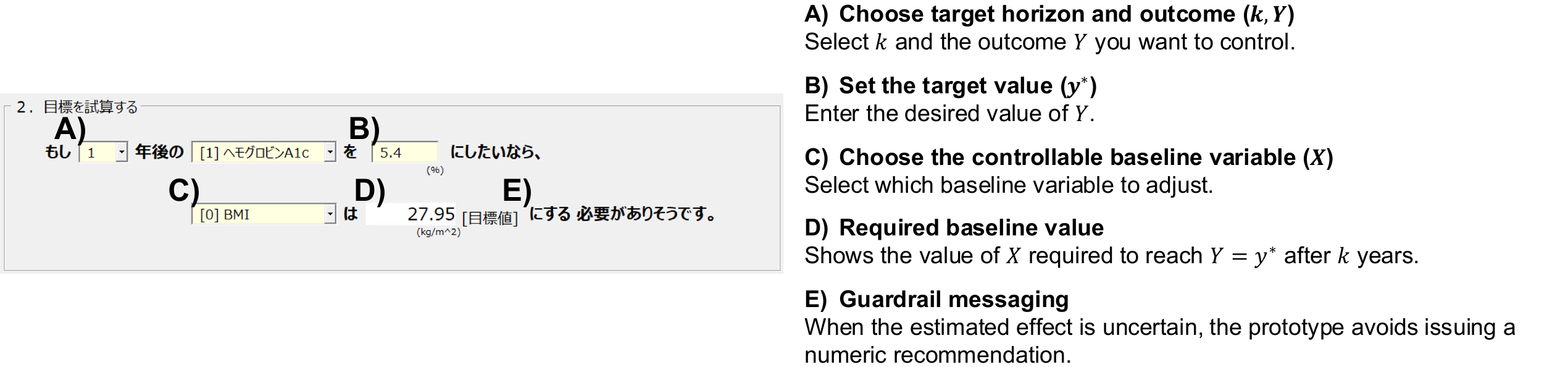}
    \caption{Goal-seeking (Target setting): specifying a desired outcome value and solving for the required baseline value under the fitted system.}
    \label{fig:proto_goal}
\end{figure}

\begin{figure}[H]
    \centering
    \includegraphics[width=\linewidth]{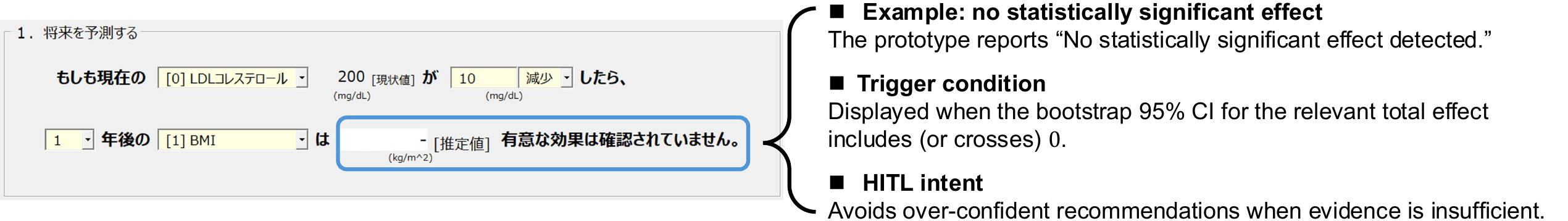}
    \caption{Decision messaging (guardrail): suppressing recommendations when the bootstrap 95\% CI includes 0.}
    \label{fig:proto_guardrail}
\end{figure}

\end{document}